\documentclass[twocolumn]{aastex631}

\usepackage{amsmath}
\usepackage{float}

\begin{document}

\title{Cosmic Ray Diffusion in the Turbulent Interstellar Medium:\\ Effects of Mirror Diffusion and Pitch Angle Scattering}

\author[0000-0003-2164-9055]{Lucas Barreto-Mota}
\affiliation{Instituto de Astronomia, Geofísica e Ciências Atmosféricas da USP,\\Departamento de Astronomia, R. do Matão, 1226 - 05508-090 - São Paulo, Brazil; \\lucas.barreto.santos@usp.br  dalpino@iag.usp.br}

\author[0000-0001-8058-4752]{Elisabete M. de Gouveia Dal Pino}
\affiliation{Instituto de Astronomia, Geofísica e Ciências Atmosféricas da USP,\\Departamento de Astronomia, R. do Matão, 1226 - 05508-090 - São Paulo, Brazil; \\lucas.barreto.santos@usp.br  dalpino@iag.usp.br}

\author[0000-0002-0458-7828]{Siyao Xu}
\affiliation{Department of Physics, University of Florida, 2001 Museum Rd., Gainesville, FL 32611, USA; \\xusiyao@ufl.edu}

\author{Alexandre Lazarian}
\affiliation{Department of Astronomy, University of Wisconsin, 475 North Charter Street, Madison, WI 53706, USA; \\alazarian@facstaff.wisc.edu}







\begin{abstract}

Cosmic rays (CRs) interact with turbulent magnetic fields in the intestellar medium, generating nonthermal emission. After many decades of studies, the theoretical understanding of their diffusion in the ISM continues to pose a challenge. This study numerically explores a recent prediction termed ``mirror diffusion" and its synergy with traditional diffusion mechanism based on gyroresonant scattering. Our study combines 3D MHD simulations of star-forming regions with test particle simulations to analyze CR diffusion. We demonstrate the significance of mirror diffusion in CR diffusion parallel to the magnetic field, when the mirroring condition is satisfied. Our results support the theoretical expectation that the resulting particle propagation arising from mirror diffusion in combination with much faster diffusion induced by gyroresonant scattering resembles a Levy-flight-like propagation. Our study highlights the necessity to reevaluate the diffusion coefficients traditionally adopted in the ISM based on gyroresonant scattering alone. For instance, our simulations imply a diffusion coefficient  $\sim10^{27}cm^2/s$ for particles with a few hundred TeV within regions spanning a few parsecs around the source. This estimate is in agreement with gamma-ray observations, which shows the relevance of our results for understanding of gamma-ray emission in star-forming regions.

\end{abstract}



\section{Introduction}\label{sec:intro}

A comprehensive understanding of cosmic ray diffusion is pivotal for addressing a multitude of issues, spanning from the generation of non-thermal radiation to the analysis of chemical composition in diverse astrophysical environments, including our galaxy.

Cosmic rays (CRs) undergo the interactions with magnetic fluctuations and change the direction of their trajectories as they diffuse through the magnetized interstellar medium (ISM). The amount of time these particles spend during their journey through the ISM plays a crucial role in elucidating phenomena such as propagation, acceleration, and non-thermal emission. The quasi-linear theory (QLT) \citep{1966ApJ...146..480J,2002cra..book.....S}, as a commonly used approach for 
studying  CR diffusion, analyzes diffusion by considering small perturbations superimposed on the mean magnetic field.

 Accounting for the actual properties of MHD turbulence \citep[see][]{2019tuma.book.....B} is essential for understanding the pitch-angle scattering. For instance, \cite{2002PhRvL..89B1102Y,2004ApJ...614..757Y} demonstrated that fast modes are more important compared to the other modes, i.e., Alfv\'en and slow modes, for gyroresonant scattering of CRs by turbulence injected at the scale height of the spiral galaxies. This happens in the ionized medium in most of galactic environments, even though the fast modes are subject to collisionless damping. The variations of magnetic field strength, arising, for instance, from the turbulent compressions, induce the modification of QLT and change the scattering rate \citep{2008ApJ...673..942Y,2012ApJ...758...78L,2018ApJ...868...36X}.

 However, QLT and its modifications  still face difficulties in reproducing the observation-required diffusion coefficients 
 \citep[][]{2004ApJ...616..617S,2006MNRAS.373.1195L,2008ApJ...673..942Y,2009A&A...507..589S,2016A&A...588A..73C,2020MNRAS.493.2817K}.
 The puzzling results include, for instance, the analysis of H.E.S.S data. \cite{2018ApJ...866..143H} suggests that in the vicinity of the Vela X pulsar wind nebula, CR electrons are likely to exhibit a diffusion coefficient of $\lesssim10^{28}cm^2/s$ at $10$ TeV within the inner region spanning a few tens of parsecs around the nebula. Similar estimations have been derived from HAWC observations concerning Geminga and PSR B0656+14 pulsars wind nebulae \citep{2017Sci...358..911A}.
Even more intriguing is the recent detection by LHAASO of very high-energy (VHE) emission at hundreds of TeVs, compatible with the presence of several PeV (Pevatron) sources of CRs inside our galaxy \citep{2021Natur.594...33C}. The potential sources linked to these observations include molecular cloud-supernova remnant (MC-SNR) interaction regions \citep[e.g.][]{PhysRevLett.108.051105,PhysRevLett.109.061101,2013Sci...339..807A}, pulsar wind nebula halos (also called PWN TeV halos) \citep{PhysRevD.104.123017,yan2023origin,2024NatAs.tmp...54Y}, and Young Star Clusters (YSCs) \citep{2021MNRAS.504.6096M,2021NatAs...5..465A}. 
The Cygnus cocoon, a very extended region surrounding a cluster of massive stars \citep{2021NatAs...5..465A}, is perhaps one of the most compelling examples of this class. Its VHE emission originates from an expansive area surrounding the stars, likely due to diffusion away from them and into the ISM.
Strong assumptions are often made to explain these observations. These assumptions go from almost free parametrization of the diffusion coefficient, as being some fraction of the average ISM diffusion coefficient estimated from QLT, up to the consideration of self-confinement theories \citep[e.g.,][]{10.1093/mnras/stz2089,2023arXiv230510251G}. 

A long-standing theoretical difficulty with resonant scattering is the $90^\circ$-problem \citep{1966ApJ...146..480J}. QLT falls short in {\it accounting}  for the scattering when the particle's pitch angle approaches $90^\circ$. Resonant scattering alone fails in reversing the moving direction of particles.
Earlier investigations have explored various phenomenological models of turbulence in conjunction with QLT \citep[see e.g.][]{1990JGR....9520673M,1999ApJ...520..204G,2016ApJ...830..130S}. However, these models have faced challenges from observations and simulations \citep{2002ApJ...578L.117Q,Gabici_2019}. 
Inevitably, the solutions ought to rely on thoroughly validated models of MHD turbulence. Such models can significantly alter pitch angle scattering, acceleration, and spatial diffusion of CRs compared to previous ad hoc approaches
 \citep[e.g.][]{2002PhRvL..89B1102Y,2004ApJ...614..757Y,2007MNRAS.378..245B,2011ApJ...728...60B,2013ApJ...779..140X,2016A&A...588A..73C,2021ApJ...908..193M,2021ApJ...912..109K,2021MNRAS.502.5821F,2023MNRAS.525.4985K,2023JPlPh..89e1701L,2023ApJ...952..168M,2024ApJ...961...80G}.

The super-diffusive behavior of turbulent magnetic field lines \citep{1999ApJ...517..700L,2013ApJ...767L..39B,2013Natur.497..466E,2013ApJ...779..140X,2014ApJ...784...38L,2022MNRAS.512.2111H,2022ApJ...926...94M,2023ApJ...959L...8Z} and particle  interaction with magnetic mirrors \citep{1969ApJ...156..445K,2020ApJ...894...63X} have been combined in \cite{2021ApJ...923...53L} (henceforth LX21) to predict a new non-resonant diffusion mechanism which was termed ``mirror diffusion". LX21 demonstrated that in the presence of superdiffusion of magnetic field lines, the probability of CRs getting trapped and bouncing back and forth between the same magnetic mirrors is negligibly small. CRs do not retrace their trajectories and bounce every time from different mirrors. 
As the mirroring causes particles to reverse their moving directions, the mirror diffusion naturally solves the 90 degree problem and is a much slower diffusion process compared to the diffusion induced by gyroresonant scattering alone.

Naturally, this new process affects only the particles with Larmor radius smaller than the size of the magnetic mirrors and a pitch angle larger than the critical angle $\theta_{pitch,c}$. This is equivalent to say that $\mu=\cos \theta_{pitch}$ is less than some critical value at the balance between mirroring and scattering 
which depends on the 
amplitude of magnetic compressions and the efficiency of pitch-angle scattering (LX21).








The magnetic compressions induced by the cascade of fast and slow modes form a hierarchy of amplitudes from large compressions $\delta B_L$ at the injection scale $L$ to the smallest ones $\delta B_d$ at the dissipation scale $l_d$. These magnetic compressions induce the hierarchy of magnetic mirrors of different sizes. CRs with the largest $\mu=\mu_c$ bounce from large-scale magnetic compressions, while CRs with smaller $\mu$ bounce from magnetic compressions at smaller scales.
As mirror diffusion preserves the first adiabatic invariant $\gamma m v^2(1-\mu^2)/2|\textbf{B}|$ \citep{1969ApJ...156..445K}, the initial value of $\mu$ does not stochastically change, but changes between $\pm\mu$, and thus, the diffusion coefficient for mirror diffusion alone is a function of $\mu$.
Effective mirroring requires that the CR gyroradius is smaller than the parallel extension of the compressible magnetic fluctuation. As $\mu_c$ where the transition between mirror diffusion and scattering diffusion takes place depends on CR energy (LX21), thus, the overall CR diffusion depends on its Larmor radius, i.e., on its energy, but with the energy dependence different from the scattering diffusion alone\footnote{Note that mirror diffusion and scattering diffusion are two separate mechanisms. Calling it ``mirror scattering" would be conceptually incorrect.}.

In highly compressible environments like molecular clouds, we expect fast modes to dominate the mirroring effect \citep{2020ApJ...894...63X,2021ApJ...922..264X}. The resulting diffusion coefficient for mirror diffusion shows a strong $\mu$-dependence \citep[LX21 and][]{2023ApJ...959L...8Z}. 
The dependence on $\mu$, makes the mirror diffusion very different from the classical resonance scattering diffusion that isotropizes $\mu$. If we are interested in the total CR diffusion flux, we need information on the distribution of $\mu$, a new element for the CR propagation paradigm. It is reasonable to assume that CRs have the isotropic $\mu$-distribution near the source. With this assumption, considering mirror diffusion, \cite{2021ApJ...922..264X} estimated the spectrum of CRs in molecular clouds (MCs)  with highly compressible turbulence subject to collisions with supernova remnants (SNR) and found that it leads to a steeper high-energy gamma-ray spectrum which can naturally explain observed systems like IC 443 and W44.

Certain analytical predictions from LX21 were also validated successfully in \cite{2023ApJ...959L...8Z}. However, the crucial importance of magnetic bouncing/mirroring and the resulting mirror diffusion motivates us for more detailed studies of the processes. In this work, we investigate the mirror diffusion of charged CR particles in a highly compressible environment using 3D-MHD simulations of an MC-like environment as the background and test particle propagation models to evaluate the diffusion behavior of the particles aiming at testing numerically the aforementioned predictions of LX21.

This paper is structured as follows. In Section \ref{sec:methodology}, we describe our model and the setup used in our simulations, and in Section \ref{sec:results} we present the results of our test particle simulations. In Section \ref{sec:discussion} we discuss our results and compare them with theoretical predictions and observations, and  in Section \ref{sec:conclusion}, we draw our conclusions.

\section{Simulation setup}\label{sec:methodology}

\subsection{Background MHD turbulence}\label{subsec:background_mag}

In order to obtain a fully developed turbulent magnetic field required to study diffusion of test particles, we employ a 3D-MHD model of a molecular cloud produced in \citet{2021MNRAS.503.5425B}. This model has a sonic Mach number $M_s=7.0$, an Alfvénic Mach number $M_A=0.6$ with a mean magnetic field parallel to the z direction, and a cubic box domain with periodic boundaries and a resolution of $512^3$. The model is isothermal and includes self-gravity in the later stages, but in this study, we consider an earlier snapshot of the simulation before self-gravity is turned-on, but with turbulence actively present and the cascade fully developed. Turbulence is driven solenoidally in the Fourier space,  around a characteristic wavelength that defines the injection scale. In this particular model, the injection scale is set to $1/4L_0$, where $L_0$ corresponds to the box size. 

\begin{figure}[ht]
\begin{center}
 \includegraphics[width=0.95\columnwidth]{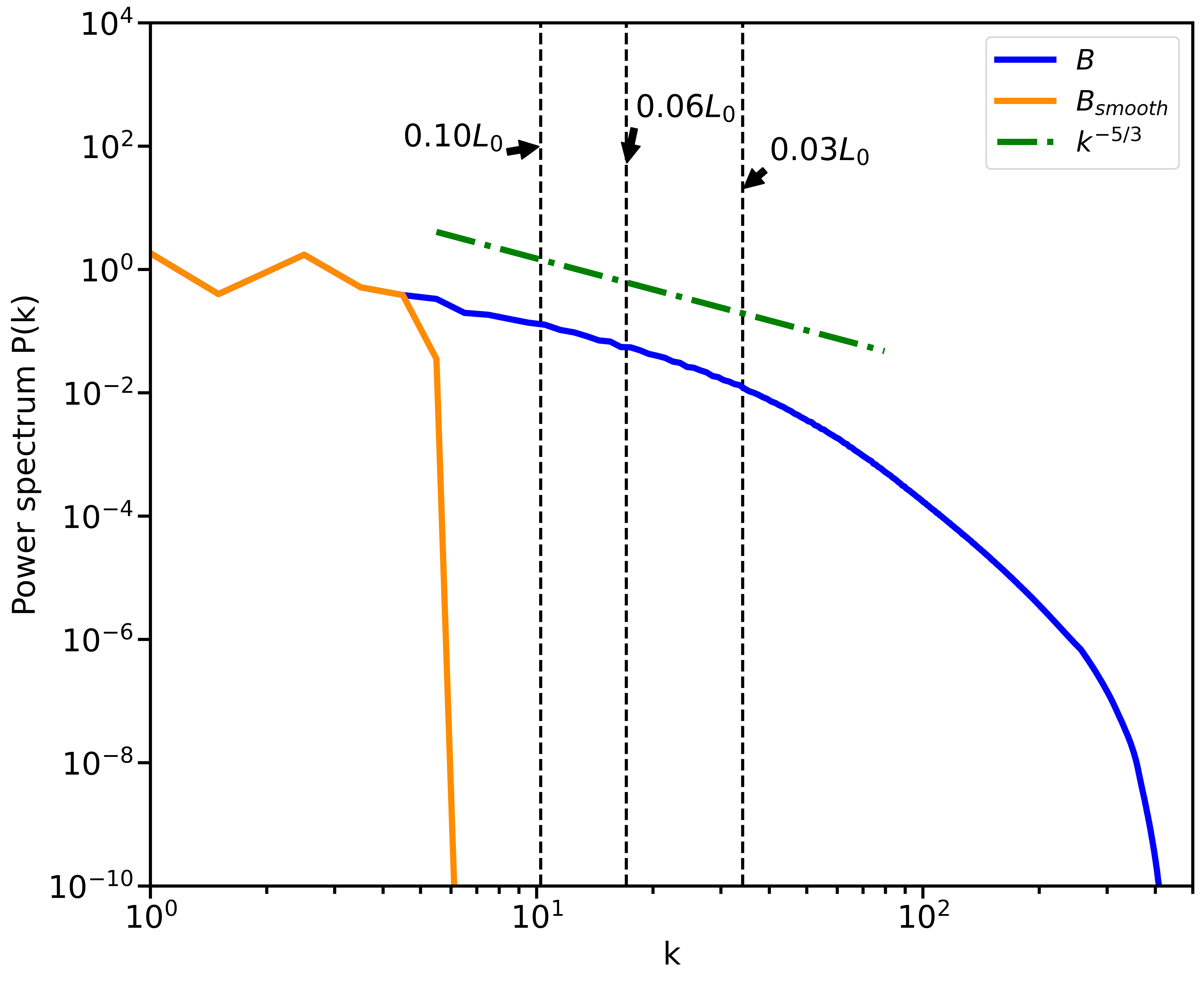}
    \caption{Magnetic power spectrum of the magnetic field components (solid lines). 
    The blue line shows the result for the original complete magnetic field background (complete models as referred in the text). The orange line shows the same for the smoothed magnetic field background (see text for details). For reference, the vertical dotted lines show the corresponding value of k for each of the particle Larmor radius considered in this work.}
    \label{fig_powerspec_B}
\end{center}
\end{figure}

Figure \ref{fig_powerspec_B} shows the power spectrum of the turbulent magnetic field for the background MHD model, after turbulence has reached a nearly steady state \citep[see][for details]{2021MNRAS.503.5425B}. The blue line shows the result for the complete background magnetic field while the orange one shows the same model with a smoothed background magnetic field, where the smaller scale perturbations have been suppressed. To smooth the small scale perturbations, we calculated the 3D Fourier transform of each component of the magnetic field, set to zero the values corresponding to $k>5$, and then  transformed it back to physical space. The resulting spectrum vanishes for $k>5$ (orange curve). From now on, we refer to this value as $k_{smooth}$. The green dashed line shows the Kolmogorov scaling with $k^{-5/3}$ for comparison. We see that a substantial part of the inertial range of the magnetic field in this smoothed model is eliminated. 

In this study, we will employ both the smoothed and the complete background turbulent magnetic field models for the CR propagation analysis and compare their results. We note that  smoothing  the magnetic field significantly reduces the effects of resonant scattering on the propagation of particles with their gyroradii smaller than the smoothing scale, enabling  us to highlight the effects of mirror diffusion. 
In the case when the particle Larmor radius $R_L$ is smaller than the damping scale of MHD turbulence, i.e., smoothed case, resonant scattering must be severely affected as the resonant condition is not satisfied, but non-resonant mirroring by larger-scale
turbulent magnetic mirrors can still happen.

{}

\subsection{Test Particle Simulations}\label{sec:testpartsim}

To study the CR diffusion in the turbulent magnetic field generated in the aforementioned model, we employ the test particle code developed by \cite{2011ApJ...728...60B} \citep[see also][]{2013ApJ...779..140X,2022MNRAS.512.2111H}. 
The particles are allowed to diffuse through the statistically steady-state background turbulent magnetic field of the simulated model described in Section \ref{subsec:background_mag}.

To follow the propagation of the particles through the turbulent magnetic fields and derive their trajectories, we compute the Lorentz force in each position along the particle trajectory:
\begin{eqnarray}
    \frac{d\mathbf{v}}{dt} = \frac{q}{\gamma mc} \mathbf{v}\times \mathbf{B},
\end{eqnarray}
\noindent where $\mathbf{v}$ is the particle velocity, $\gamma$ its Lorentz factor $\gamma = 1/\sqrt{1 - v^2/c^2}$, and $\mathbf{B}$ is the magnetic field.

Each charged particle has:

\begin{eqnarray}
     \label{eq:larmor_radius}R_L &=& \frac{v_\perp}{\Omega}= \frac{\gamma mc v_\perp}{q B} 
\end{eqnarray}

\noindent where $R_L$ is the Larmor radius, $m$ the mass, $q$ the charge, and $\Omega$ the gyrofrequency. In this study, we consider the particle velocity given by the light speed $c$. 
We take  particles with constant Larmor radius and initial  $v_{\perp,0}\sim c$.

The code uses a Burlisch-Stoer method to trace the trajectories of test particles and uses an adaptive time step. At each time step, the magnetic field at the grid points is interpolated to the test particles position using a cubic spline routine\footnote{The code uses $10^3$ neighboring points for the cubic spline interpolation. The large number of points allows for higher accuracy and smoother derivatives.} \citep{1986nras.book.....P,2013ApJ...779..140X}.

Given that the MHD turbulent models are scale-free and the energy of the particles depends on the properties of the simulated box (see Section \ref{sec:discussion}), one can scale the system knowing the size of the domain $L_0$, the mean density and the sound speed as well as the scale independent parameters $M_A$ and $M_s$ that are defined at the injection scale of turbulence.

\subsection{Initial setup}\label{subsec:initial_cond}

The test particle models we simulated are summarized in Table \ref{table:all_models}.

\begin{table}[!ht]
\begin{center}
\caption{Initial conditions for all models.}
\begin{tabular}{cccc}
\hline
Models & $\mu_0=\cos\theta_0$ & $R_L/L_0$ & $k_{smooth}$ \\ \hline \hline
r03mu20                       & 0.2                & 0.03    &           \\
r03mu80                       & 0.8                & 0.03    &           \\
r06mu20                       & 0.2                & 0.06    &           \\
r06mu80                       & 0.8                & 0.06    &           \\
r10mu20                       & 0.2                & 0.10    &           \\
r10mu80                       & 0.8                & 0.10    &           \\
r03mu20k5                     & 0.2                & 0.03    & 5         \\
r03mu80k5                     & 0.8                & 0.03    & 5         \\
r06mu20k5                     & 0.2                & 0.06    & 5         \\
r06mu80k5                     & 0.8                & 0.06    & 5         \\
r10mu20k5                     & 0.2                & 0.10    & 5         \\
r10mu80k5                     & 0.8                & 0.10    & 5         \\
\hline
\end{tabular}

\label{table:all_models}
\end{center}
\end{table}

Each model is characterized by two free parameters, namely, the initial pitch angle ($\theta_0$) between the particles velocity and the local magnetic field, which we consider to be either $\mu_0 = \cos\theta_0 = 0.2$ or $ 0.8$, and the Larmor radius  ($R_L$), $R_L/L_0 = 0.03, 0.06$ and $0.10$, in terms of the number of cells of the background simulation, $R_L (\# cells)=15, 30$, and $50$, respectively (see the representation of these values in Figure \ref{fig_powerspec_B}). We note  that particles are not accelerated and their energy stays the same in our simulations, we focus only on their diffusion. Each model name is labeled with these two parameters in Table \ref{table:all_models}. As described in section \ref{subsec:background_mag}, we also consider two sets for the magnetic field configuration. One with the entire background turbulent magnetic field distribution, and another where the small-scale perturbations of the field were suppressed for $k_{smooth} = 5$, which corresponds to a dissipative scale of about $\sim 0.2L_0$ ($\sim100$ grid cells)\footnote{Values of $k_{smooth}>5$ were tested, but the results showed little to no notable differences with regard to the model with the complete distribution of the magnetic field.}. These latter models are labeled $k5$ in Table \ref{table:all_models}. The dissipation scale of the complete distribution of magnetic field is around $k\sim250$, which corresponds to $0.004L_0$ approximately ($\lesssim3$ grid cells). In all models, 3000 test particles are injected. 


\section{Results}\label{sec:results}
\subsection{Particle trajectory and mirroring identification}\label{sec:trajectory_bounceid}

\begin{figure*}[ht]
\begin{center}
 \includegraphics[width=1.99
 \columnwidth,angle=0]{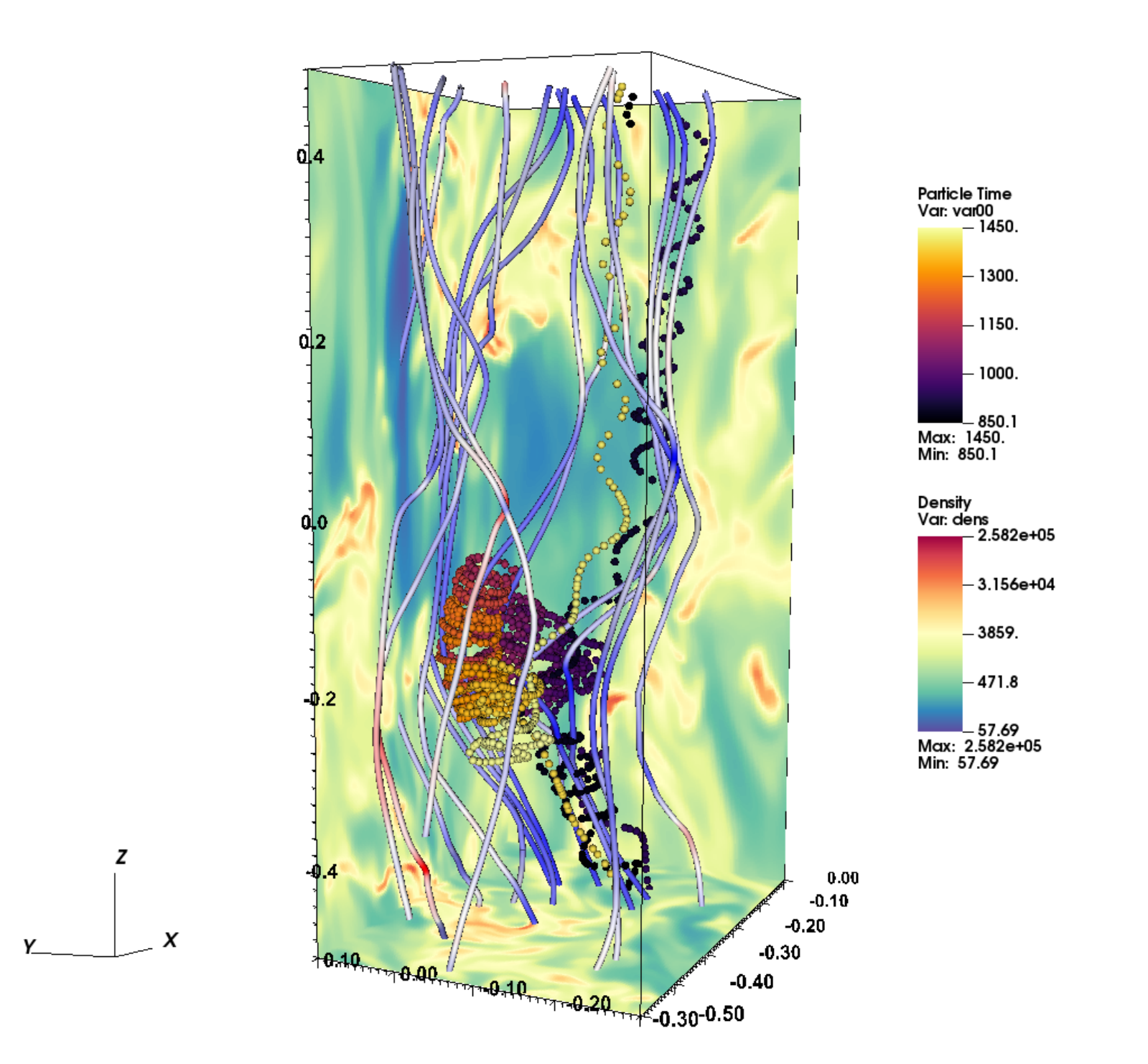}
    \caption{Example of a test particle trajectory (represented by solid colored  circles) with $R_L/L_0\simeq0.03$ and $\mu_0 = 0.2$ in the background with the complete distribution of the magnetic field.  Initially,  the particle travels upwards parallel to the magnetic field (black circles), reentering the domain from the bottom face until it encounters magnetic mirrors that trap it for a while (dark purple to orange circles), diffusing perpendicularly to the field lines until it escapes again traveling downwards and reentering through the upper region of the domain, finding another mirror at the end of its trajectory (bright yellow circles). 
    A few field lines are shown in red, white, and blue for visual distinction only. The density background is included 
    to highlight the turbulent compressions of the medium.
    The color scales on the right hand side show the time elapsed in the  particle trajectory between 850 and 1450 (same units as  in Figure \ref{fig_testPart_muXt_r15}), and the density of the background (in code units), respectively.
    See Figure \ref{fig_testPart_muXt_r15} and text for further details.  
    We have plotted only a section of the cubic box in order to make the particle trajectory clearer.
    }
    \label{fig_testPart_trajectory}
\end{center}
\end{figure*}{}

Figure \ref{fig_testPart_trajectory} depicts an example of a particle trajectory in the original MHD turbulence simulation. A characteristic behavior observed in every model is long flights interrupted by successive bounces, where the particle interacts with different magnetic mirrors (Levy-flght-like propagation). As discussed in LX21, only particles with pitch angles above a critical value should interact with magnetic mirrors present along their path (Equation \ref{eq:mucrit}). 
When these conditions are met, the particle reverses its moving direction and bounces between mirrors. 
In the meantime, the particle also undergoes superdiffusion in the direction perpendicular to the mean magnetic field as a result of the turbulent reconnection of the magnetic fields. Because of the perpendicular superdiffusion, the particle does not remain trapped by mirrors, but diffusively move along the magnetic field when interacting with different mirrors\footnote{As mentioned in Section \ref{sec:intro}, the effect of perpendicular superdiffusion has been thoroughly 
investigated elsewhere. We refer to numerical studies \citep[e.g.][]{2013ApJ...767L..39B,2013ApJ...779..140X,2013Natur.497..466E,2014ApJ...784...38L,2022MNRAS.512.2111H,2022ApJ...926...94M,2023ApJ...959L...8Z} that confirm the superdiffusive behavior of magnetic field lines and cosmic rays following the magnetic field \citep{1999ApJ...517..700L,2014ApJ...784...38L}}. When the pitch angle becomes sufficiently small due to the pitch-angle scattering, the particle can escape from mirrors and undergo the scattering diffusion. With the stochastic change of pitch angle due to the gyroresonant scattering, the particle stochastically experiences both the mirror diffusion and scattering diffusion.


\begin{figure*}[ht]
\begin{center}
    \includegraphics[width=1.95\columnwidth,angle=0]{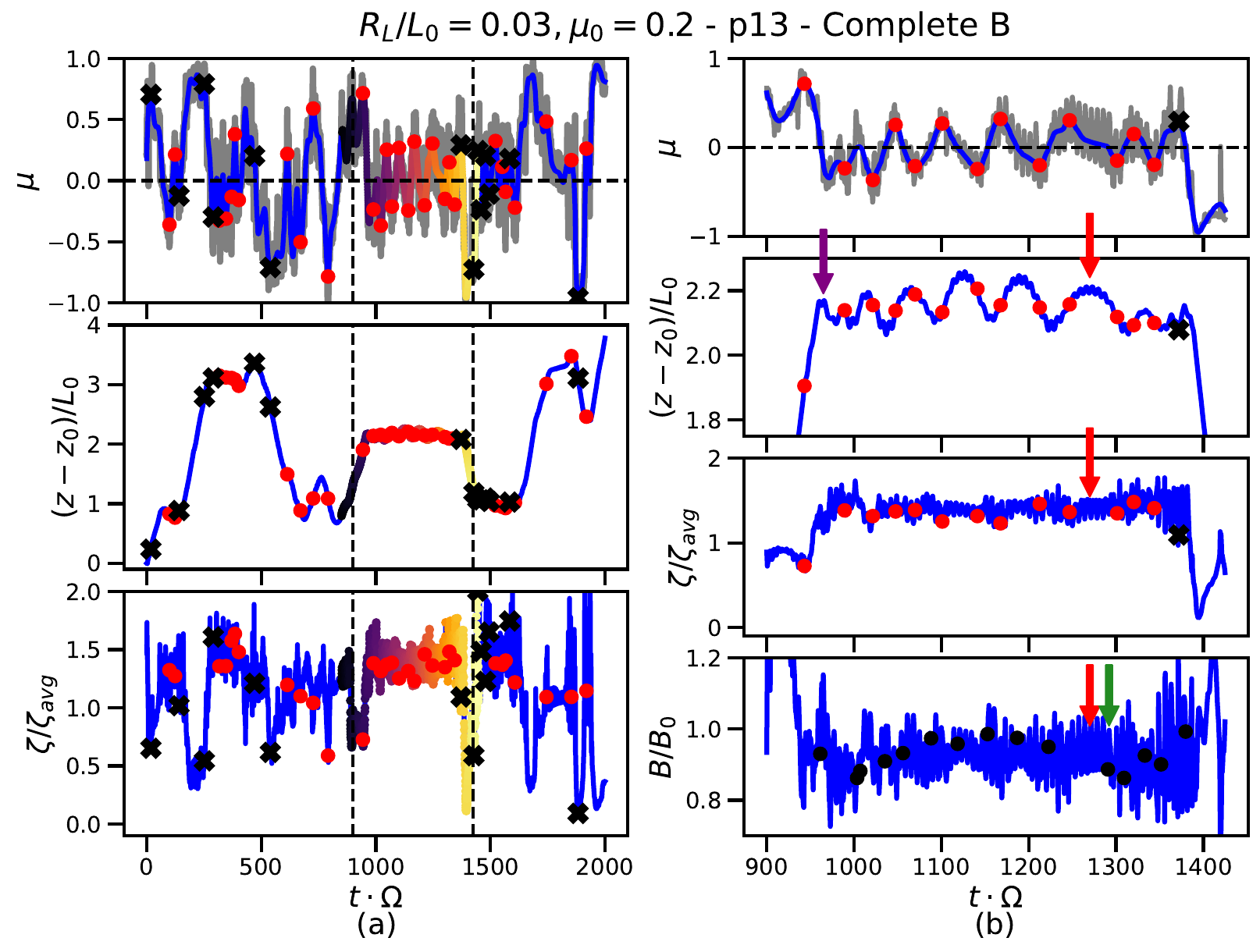}
    
    \caption{Diagrams on the left (a)  depict, for the complete background magnetic field distribution, the following quantities: top panel, the time evolution of the cosine of the pitch angle ($\mu$) for the same particle shown in Figure \ref{fig_testPart_trajectory}. 
    The gradient color scheme depicts the portion of the particle trajectory as in Figure \ref{fig_testPart_trajectory}. Gray line shows the complete variation of $\mu$ over time and the blue line shows the same, but without the fastest variations of $\mu$. The middle panel shows the displacement of the particle parallel to the magnetic field with regard to its initial position, and  the bottom panel shows $\zeta/\zeta_{avg}$ (Equation \ref{eq:adiab_inv}) over time. Red circles and black crosses are potential candidates for the beginning of bounces due to magnetic mirror interactions. Following the conservation of magnetic moment criteria introduced in the text, black crosses represent the rejected candidates.
    Time  is normalized by the initial value of the girofrequency $\Omega = c/R_L$ (Equation \ref{eq:larmor_radius}), being $c=1$ in the code units. On the right side (b), the three first panels are the same as in (a) but zoomed in with a smaller range of $t$ marked by the vertical dashed lines. The bottom panel shows the magnetic field strength normalized by the mean magnetic field along the particle trajectory and the black circles indicate where we identify $\mu=0 \, (\theta=90^\circ)$ in the top panel (see text for details).
    } 
    \label{fig_testPart_muXt_r15}
\end{center}
\end{figure*}{}

\begin{figure*}[ht]
\begin{center}
 \includegraphics[width=1.95\columnwidth,angle=0]{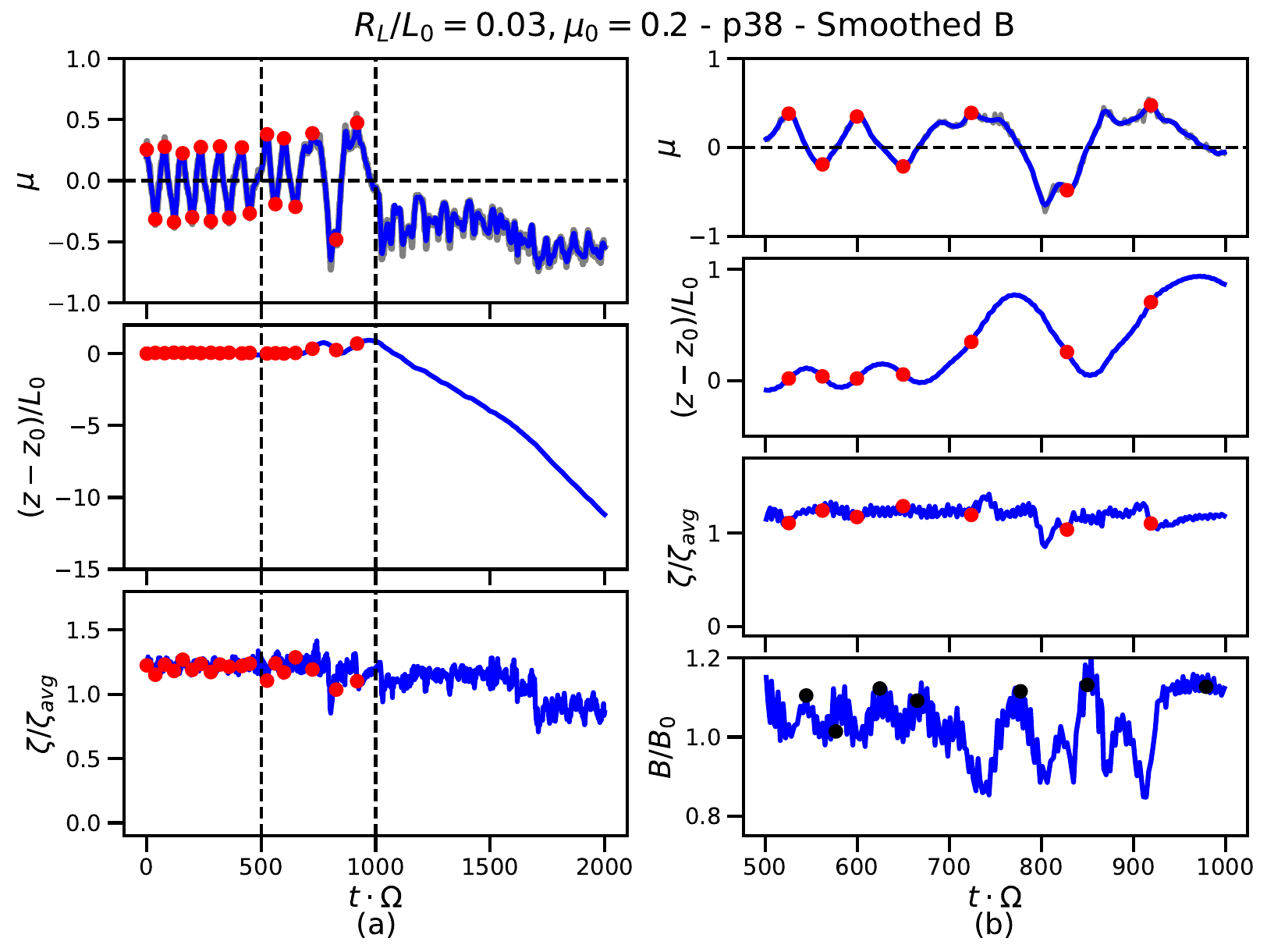}
    \caption{The same as Figure \ref{fig_testPart_muXt_r15}, but with smoothed magnetic field as background. 
    }
    \label{fig_testPart_muXt_r15_k5}
\end{center}
\end{figure*}{}

In Figure \ref{fig_testPart_muXt_r15}, we show the time evolution of $\mu$ (top panel), the distance traveled parallel to the magnetic field  versus time (middle panel), and $\zeta/\zeta_{avg}$ versus time, where $\zeta$ is the adiabatic invariant of the particle, which is conserved when the particle is bouncing in a mirror: 

\begin{eqnarray}\label{eq:adiab_inv}
    \zeta(t) = \gamma m v^2\frac{1-\mu(t)^2}{2|\textbf{B}(t)|},
\end{eqnarray}

\noindent and $\zeta_{avg}$ is the value of $\zeta$ calculated for $\mu=0.5$ and the average magnetic field $B_0$, for the same particle shown in Figure \ref{fig_testPart_trajectory}. 

As an example, Figure \ref{fig_testPart_trajectory} shows just a part of the particle's trajectory. The respective time interval is shown in Figure \ref{fig_testPart_muXt_r15} with the same color coding. Considering this time interval, the second panel of Figure \ref{fig_testPart_muXt_r15} shows that the particle travels upward until it interacts with a mirror, staying more or less at the same height until it is able to continue its trajectory downwards after several consecutive mirroring events. In Figure \ref{fig_testPart_trajectory} we see that during these several consecutive bounces, the particle continues to move perpendicular to the mean magnetic field, interacting with different mirrors. In the lower panel of Figure \ref{fig_testPart_muXt_r15}, the respective time interval also shows a nearly constant adiabatic invariant, which reinforces the fact that these are mirroring events.

The peaks of $\mu$ immediately before each bounce point (i.e., when $\mu=0$) in Figure \ref{fig_testPart_muXt_r15} mark the beginning of a bounce, $\mu_b$, where $\mu_b$ is the value of $\mu$ at these peaks. This parameter represents the angle at which the particle begins interacting with a magnetic mirror. Specifically, if $|\mu_b|>|\mu_c|$, the particle should not be subject to mirroring,
where $\mu_c=\cos{\theta_{pitch,c}}$ is the critical pitch-angle cosine threshold for mirror diffusion, as defined in Section \ref{sec:intro}. Unless otherwise noted (e.g., in Figures \ref{fig_testPart_muXt_r15} and \ref{fig_testPart_muXt_r15_k5}), all analyses use the absolute value of $\mu_b$.

For fluctuations induced by fast modes of MHD turbulence, LX21 obtained the critical value of $\mu_c$: 

\begin{eqnarray}
    \mu_c &=& \Bigg[\frac{14}{\pi} \frac{\delta B_f^2}{B_0}\Big(\frac{v}{L\Omega}\Big)^{1/2}\Bigg]^{2/11},\label{eq:mucrit}
\end{eqnarray}
\noindent where $\delta B_f$ is the perturbed magnetic field of fast modes, $B_0$ is the mean magnetic field, $v$ is the particle velocity, $L$ is the turbulence driving scale, and $\Omega$ is the particle's gyrofrequency.

On the other hand,
in the limit of incompressible turbulence $\mu_c$ will be dominated
by  fluctuations caused by pseudo-Alfvén modes:

\begin{eqnarray}
    \mu_{c,inc} &=& \sqrt{\frac{\delta B_s}{B_0+\delta B_s}},
\end{eqnarray}

\noindent where $\delta B_s$ is the perturbed magnetic field of slow modes. 
A thorough discussion of all potential scenarios regarding 
MHD modes dominance over the bounce diffusion process is beyond the scope of this work, but see Section 5.2 of LX21 for a more detailed analysis.

To identify the peaks in $\mu$ in Figure \ref{fig_testPart_muXt_r15}, we first filter the values of $\mu$ along time using a Butterworth digital filter \citep{2020SciPy-NMeth}, which helps us to follow the average behavior of $\mu$ along time. Note that there is a fast variation of $\mu$ and that not all bounce points correspond to $\zeta/\zeta_{avg}$ constant.
Considering the smaller segment of the particle's trajectory depicted in  the (b) panels on the right side of Figure \ref{fig_testPart_muXt_r15}, where the bounces are better spaced and vary roughly between $-\mu$ and $+\mu$, we find that $\mu=0$ also corresponds to maximum magnetic field $B/B_0$ values (see bottom panel). However, if we observe the region identified with a red arrow in the (b) panels, there is a change of direction that does not correspond to $\mu=0$. Due to data filtering in the top panel, we only find $\mu=0$ where the green arrow points in the bottom panel. This misidentification occurs because filtering $\mu$ obscures some mirror interactions. The same issue arises in the region indicated by the purple arrow in the second (b) panel from top.

This variation is reduced comparing the first panels of Figures \ref{fig_testPart_muXt_r15} and \ref{fig_testPart_muXt_r15_k5}. Figure \ref{fig_testPart_muXt_r15_k5} shows the same information presented in Figure \ref{fig_testPart_muXt_r15}, but for a particle propagating in the smoothed background distribution of the magnetic field. In Figure \ref{fig_testPart_muXt_r15}, top panels, the gray line (original data) has a component that partially hides the average behavior of the particle which is depicted by the blue line (filtered data). In Figure \ref{fig_testPart_muXt_r15_k5}, since the magnetic field background is smoothed and scattering is already reduced, the peaks are more easily identifiable. Not only that, but the changes in the particle direction clearly correspond to maximum values  of $B/B_0$ in Figure \ref{fig_testPart_muXt_r15_k5}(b).
In summary, in  the smoothed case, identifying the bounces is much easier because scattering at small scales is not involved and also there are fewer mirror structures at smaller scales.   Nevertheless, a similar trend is observed in both the complete and smoothed cases, with $B/B_0$  maximums generally corresponding to $\mu=0$, as expected for mirroring.

In turbulence, particle trajectories are distorted by scattering. To distinguish such events from mirror diffusion, we introduce the following criterion. Let us recall that $\mu_b$ is the peak value of $\mu$  right before a  bounce.  We can thus evaluate the respective adiabatic invariant of $\mu_b$, $\zeta$ using Equation \ref{eq:adiab_inv}. This is shown in the third  panels from top of Figures \ref{fig_testPart_muXt_r15}  and \ref{fig_testPart_muXt_r15_k5}. For a particle with $n$ occurrences of $\mu=0$, and each peak occurring at times $t_1<t_2<...<t_i<...<t_n$, we compare $\zeta(t_i)/\zeta_{avg}$ with $\zeta(t_{i-1})/\zeta_{avg}$ and $\zeta(t_{i+1})/\zeta_{avg}$ (to simplify notation, we consider $\zeta_i=\zeta(t_i)/\zeta_{avg}$). With this, for a given bounce at time $t_i$, $\mu_b(t_i)$ is discarded if $|\zeta_i - \zeta_{i-1}|>0.25$ and $|\zeta_i - \zeta_{i+1}|>0.25$. In other words, if the variation of $\zeta_i$ is bigger than $0.25$ compared to its neighbors, the bounce is discarded\footnote{We note that the small oscillations in magnetic moment (as well as in $\mu$ and in $B$) are associated with gyrations, which are fundamentally different from the large stochastic variations caused by  scattering. The threshold value $0.25$ was chosen to distinguish between these two effects.}.
For the first identified point, $\zeta_1$, we compare with the values of the next two identified points whilst for the last one, $\zeta_n$, we compare with the two before. If there are less than 3 bounce events, we simply accept the values of identified $\mu_b$.

For $R_L/L_0 = 0.03$ in the complete magnetic field background, over $73\%$ of bounces satisfy the condition defined above, $|(\zeta_i - \zeta_{i-1})/\zeta_{avg}|<0.25$. This proportion decreases slightly to $68\%$ and $67\%$,
for $R_L/L_0 = 0.06$ and $0.10$, respectively. When the condition is tightened to $|(\zeta_i - \zeta_{i-1})/\zeta_{avg}|<0.2$, 
the corresponding percentages drop to $66\%$, $60\%$ and $59\%$ for $R_L/L_0 = 0.03$, $0.06$ and $0.10$, respectively. These results 
indicate the important role of magnetic mirroring in reducing the diffusion of particles. This point will be further explored in the following sections.

We have also tested an alternative method to identify bounces, which is described in Appendix \ref{appendix:alt_mirror_criteria}. We find that the results align with  those described in this and the next section.

\subsection{Pitch angle distribution of $\mu_b$}\label{sec:ang_distribution}

To evaluate the pitch angle distribution of bounce events, we will use the values obtained of $\mu_b$ and compare with the maximum critical value (LX21):

\begin{eqnarray}\label{eq:muc_max}
    \mu_{c,max} \simeq \sqrt{\frac{\delta B}{\delta B + B_0}},
\end{eqnarray}

\noindent where $\delta B$ is the perturbed field locally defined by the position of the particle and $B_0$ is the mean magnetic field\footnote{We note that this equation corresponds to the maximum value of $\mu_c$ which is attained when $\mu_c$ becomes independent of the particle energy (LX21).}.

\begin{figure}[ht]
\begin{center}
 \includegraphics[width=1.0
 \columnwidth,angle=0]{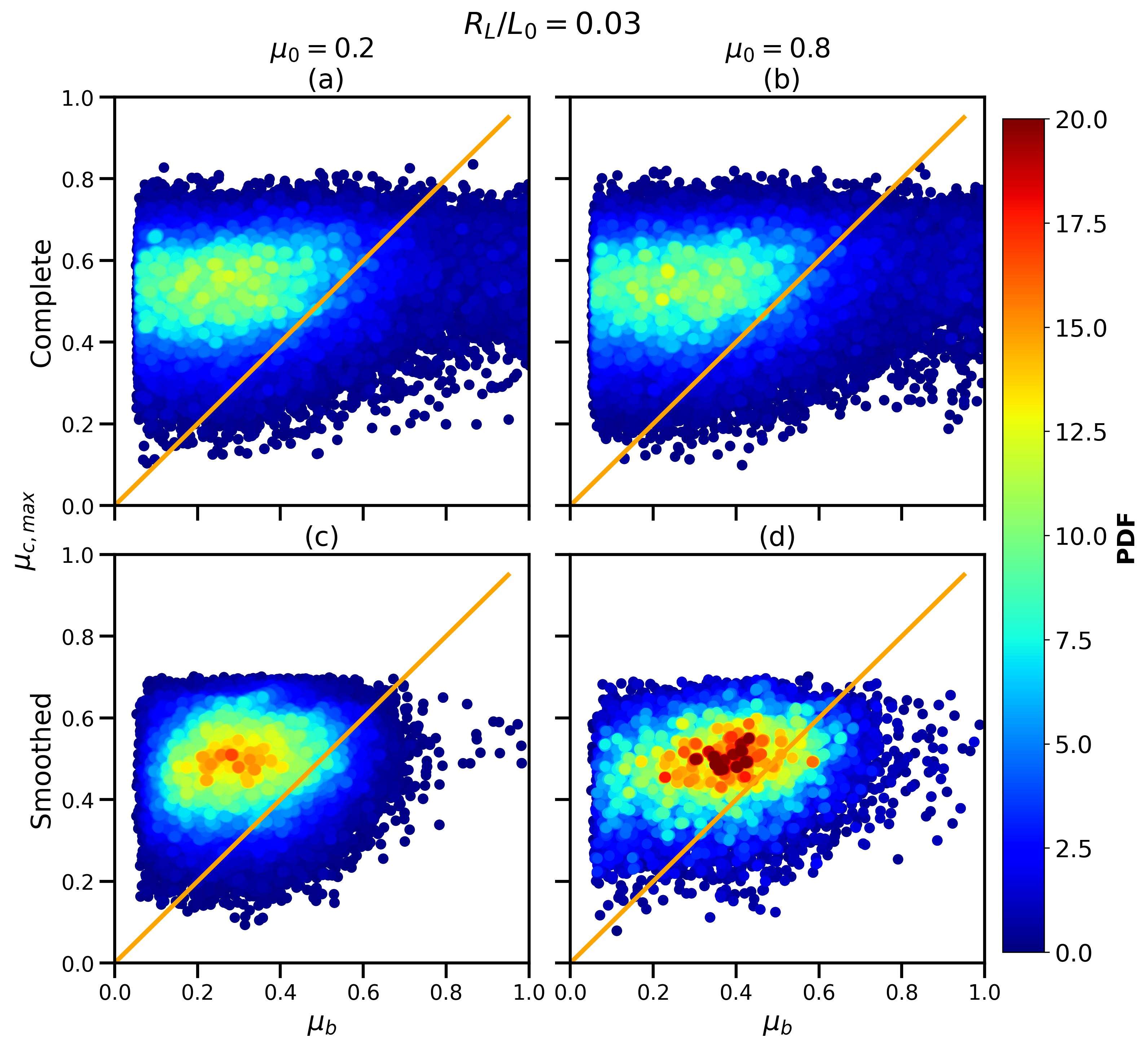}
    \caption{Comparison between the expected values of $\mu_{c,max}$ (Equation \ref{eq:muc_max}) and the identified bounce values from the simulations $\mu_b$. The orange line indicates where $\mu_{c,max} = \mu_{b}$. The color maps show the 2D probability density function normalized such that the sum over a bin value times the bin area is equal to 1. Only particles with more than two mirror interactions have been included.
    Top panels show the results for models with the complete magnetic field distribution, while the bottom panels show the results for models with the smoothed distribution of magnetic fields. On the left hand side we show models with $\mu_0=0.2$, while on the right hand side we show models with $\mu_0 = 0.8$.}
    \label{fig_mucXmub}
\end{center}
\end{figure}{}

\begin{figure}[ht]
\begin{center}
 \includegraphics[width=1.0
 \columnwidth,angle=0]{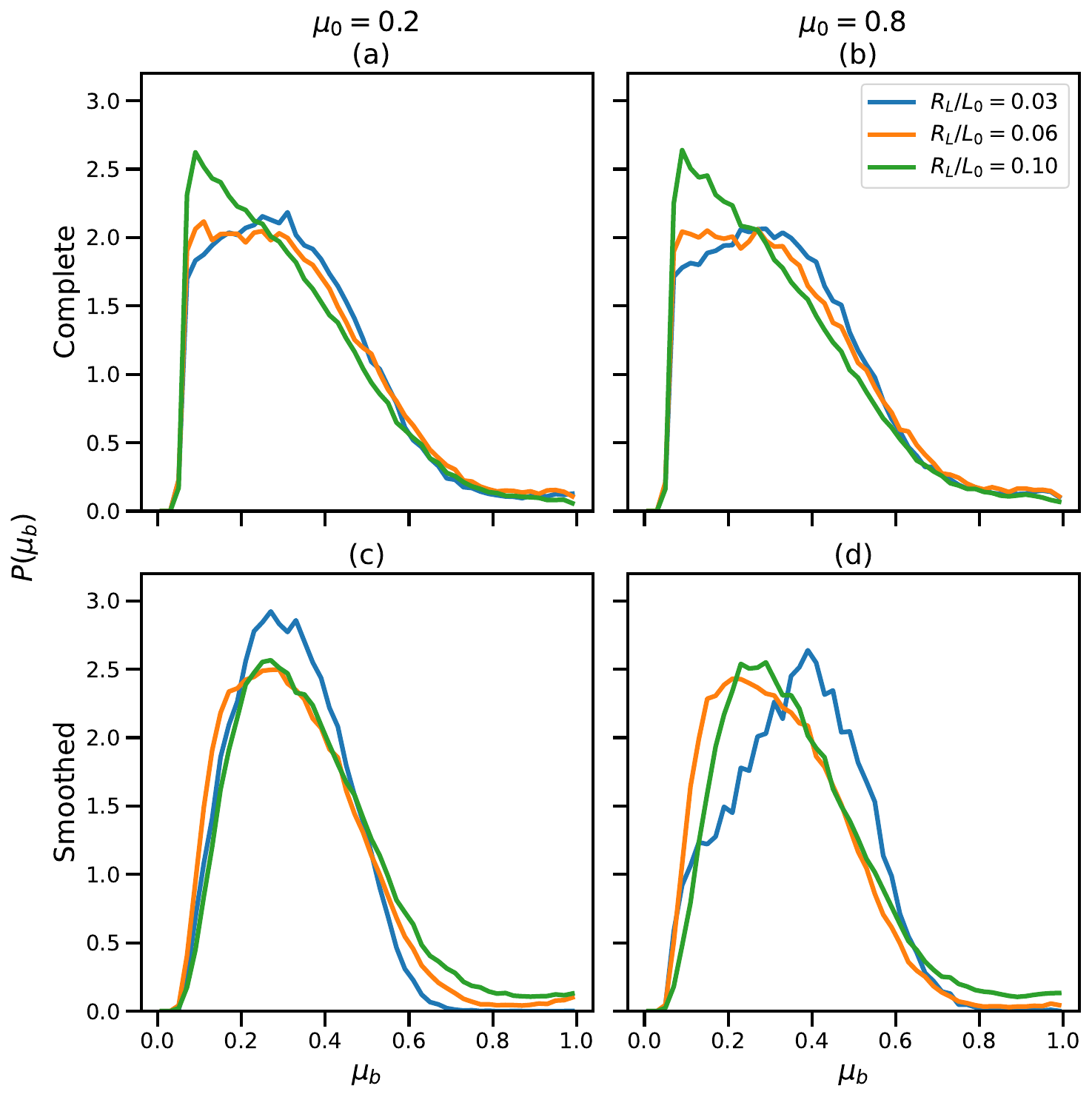}
    \caption{Probability density distributions of $\mu_b$ (see text for further details). Colors indicate different Larmor radii of the particles. The top panels show the PDFs for models with the complete background magnetic field, whilst the bottom panels show PDFs for models with the smoothed magnetic field as background. On the left hand side we show models with $\mu_0=0.2$, while on the right hand side we show models with $\mu_0 = 0.8$. The distributions are independent of each other. The variance in the complete background model stays between 0.03 and 0.04. For the smoothed case stays between 0.02 and 0.03, except for the r15mu20k5 which has a variance closer to 0.01.}
    \label{fig_PDF_bounceMu}
\end{center}
\end{figure}{}

Considering what was discussed in Section \ref{sec:trajectory_bounceid} and the information in Figures \ref{fig_testPart_muXt_r15} and \ref{fig_testPart_muXt_r15_k5}, we can produce the distribution of $\mu_b$ and its probability density distributions (PDFs) for all models.

Figure \ref{fig_mucXmub} compares the values of $\mu_b$ with the respective $\mu_{c,max}$ for the bounces identified in the models with $R_L=0.03L_0$. The color maps show the 2D probability density function normalized such that the sum over a bin value times the bin area is equal to 1. The top panels show the results for models with complete distribution of magnetic fields, while the bottom panels show the same for models that consider the smoothed distribution of magnetic fields.
When $\mu_0$ is small, the number of bounces is generally higher than when $\mu_0$ is closer to 1. When the upper and bottom panels are compared, it is clear that the lack of scattering plays an important role in slowing down the diffusion parallel to the magnetic field, as evidenced by the models with a smoothed background.

In the smoothed models in Figure \ref{fig_mucXmub} (panels (c) and (d)), we see that the identified $\mu_b$ events are mostly smaller than the corresponding predicted maximum values $\mu_{c,max}$. About $80-85\%$ of all the identified mirror events are above the orange line for all panels in Figure \ref{fig_mucXmub}. This means that the particles are bouncing from larger-amplitude mirrors in the MHD cascade. As shown in Figure \ref{fig_testPart_muXt_r15_k5}, the peaks maintain similar values while the particle bounces through successive mirrors. 
In addition, $\zeta$ stays almost constant during this process, indicating that $\mu_b$ is a good proxy for the bounce process. Note that for panel (d), particles with less than 3 mirror interactions are not included.

This scenario is quite different when scattering is more important (panels (a) and (b) of Figure \ref{fig_mucXmub}). The initial pitch angle is less important for the overall behavior of the test particles and the identification of bounces with constant $\zeta$ is harder due to the presence of variations that hide the average behavior of the particle,
introduced by pitch angle scattering. Comparing $\mu_b$ with its respective $\mu_{c,max}$ for this case shows a larger number of points below the orange line in Figure \ref{fig_mucXmub}. Points identified here as bounce have already been compared with their neighbors, according to what was discussed in Section \ref{sec:trajectory_bounceid}. These deviations can generally be explained by one of two reasons: particles are scattered and change direction, or due to scattering, $\mu$ later becomes smaller than $\mu_{c,max}$ along the particle trajectory, and then the particle bounces while interacting with a mirror. In the former case, misidentifying occurs because our criterion is not strict enough. Still, the number of mirroring events is much more numerous than the events of bounces caused by scattering. In the latter case, we are not able to track the beginning of a bounce by looking at the peaks of $\mu$ versus time, so that defining a $\mu_b$ for these cases can be somewhat arbitrary. Still, these events are not frequent enough to interfere in our analysis\footnote{We stress that mirror diffusion is  present in both cases. If resonant scattering is weak, mirror diffusion will still act over the particles and can reduce the diffusion parallel to the magnetic field. In the smoothed case we also remove mirrors associated to small scales, which allows particles to travel further between mirrors.}. 

The statistical significance of the region above the orange (critical) line in the plots of Figure \ref{fig_mucXmub} is clear.
This line serves as a guide. As mentioned earlier, the identification of mirror interactions and scatterings in the turbulent magnetic fields is complex and makes it difficult to define where one effect ends and another begins. Nonetheless, the fact that most bounces occur with a nearly constant adiabatic invariant (Figures \ref{fig_testPart_muXt_r15} and \ref{fig_testPart_muXt_r15_k5}) and follow the maximum critical value for the interaction suggests that mirror diffusion is a significant effect and aligns with LX21 predictions.

Figure \ref{fig_PDF_bounceMu} shows the PDFs of $\mu_b$ for all models. We calculate the PDF for all bounces integrated in time, after $t\times\Omega=4000$. Different colors represent different initial Larmor radii for the particles, with panels on the left-hand side showing models with an initial pitch angle such that $\mu_0 = 0.2$ and on the right-hand side with $\mu_0 = 0.8$. The top two panels show the results for models with the complete magnetic field whilst the bottom panels show results for models with the smoothed field as the background.

In these PDF distributions all models exhibit a drop in probability for $\mu_b \gtrsim 0.6$, along with a prominent peak between $0.2$ and $0.3$ in most cases. However, there is a noticeable scarcity of bounces identified with $\mu_b$ close to zero—not because such events are absent (see Figure \ref{fig_testPart_muXt_r15}), but because they are filtered out in the complete background models, where they appear as rapid fluctuations in $\mu$. In the smoothed background models, these near-zero $\mu_b$ bounces occur far less frequently. There is a correlation between $\mu$ and mirror size, as shown in Figure 6 in \cite{2023ApJ...959L...8Z}. Small $\mu$ particles mainly interact with smaller mirrors, as they are more efficient in mirroring with larger longitudinal magnetic field gradient. 
In the smoothed case, where small mirrors are absent, the mirroring of small $\mu$  particles by larger mirrors is less efficient than in the complete case.
  
For the smoothed model \textit{r03mu80k5}, only $\sim50\%$ of the particles interact with a magnetic mirror, 
and about $1/5$ of the bouncing particles suffer only one or two bounces along the trajectory\footnote{We note that in this case, due to the absence of at least three bouncing events, these instances have been discarded from the analysis.}. This introduces two points that need to be considered. First, 
the fraction of mirroring particles in this model is much smaller than in other models, where the pitch angle of the particle either is smaller at the beginning or resonant scattering changes the particle's pitch angle enough that their probability of interacting with a mirror is higher.
Second, as the particles travel inside the domain with higher values of $\mu$, they will see changes in the magnetic field and will only interact when the conditions for bounce are met, which will be skewed towards higher $\mu$ values due to the initial condition. 
The probability of a mirroring event occurring with $\mu>0.6$ is still small, but there are further complications that needs to be considered as we will discuss in the next section.

To complete the discussion in this section, in Table \ref{tab:mirror-events} we compare the total number of identified bounce events  with the total number of candidates for all cases in Figure \ref{fig_PDF_bounceMu}. Clearly, the percentage is larger for the smoothed models and for smaller particle Larmor radius. See also the discussion in Appendix \ref{appendix:alt_mirror_criteria} on how these percentages increase for an alternative method of identification of the bounce events. Nonetheless, the overall results remain consistent with the findings discussed above.

\begin{table*}[]
\centering
\caption{Comparison between mirror events and the total number of candidates.}
\label{tab:mirror-events}
\begin{tabular}{cccccccc}
\hline
Complete & $\mu_0=0.2$   &        &       &  & $\mu_0=0.8$   &        &       \\ \hline
$R_L$    & Mirror Bounce & Total  & \%    &  & Mirror Bounce & Total  & \%    \\ \hline
0.03     & 82636         & 123859 & 66.72 &  & 79733         & 120331 & 66,26 \\
0.06     & 134473        & 223060 & 60.29 &  & 131082        & 218128 & 60.09 \\
0.10     & 190035        & 322056 & 59.01 &  & 181022        & 307357 & 58.90 \\ \hline\hline
Smooth   & $\mu_0=0.2$   &        &       &  & $\mu_0=0.8$   &        &       \\ \hline
$R_L$    & Mirror Bounce & Total  & \%    &  & Mirror Bounce & Total  & \%    \\ \hline
0.03     & 265178        & 354073 & 74.89 &  & 114223        & 152117 & 75.09 \\
0.06     & 113829        & 154984 & 73.45 &  & 102410        & 136214 & 75.18 \\
0.10     & 90334         & 140372 & 64.35 &  & 87864         & 137993 & 63.67 \\ \hline
\end{tabular}
\end{table*}

\subsection{Particle diffusion in MHD turbulence}\label{sec:diff_comp_models}

\begin{figure*}[ht]
\begin{center}
 \includegraphics[width=1.9
 \columnwidth,angle=0]{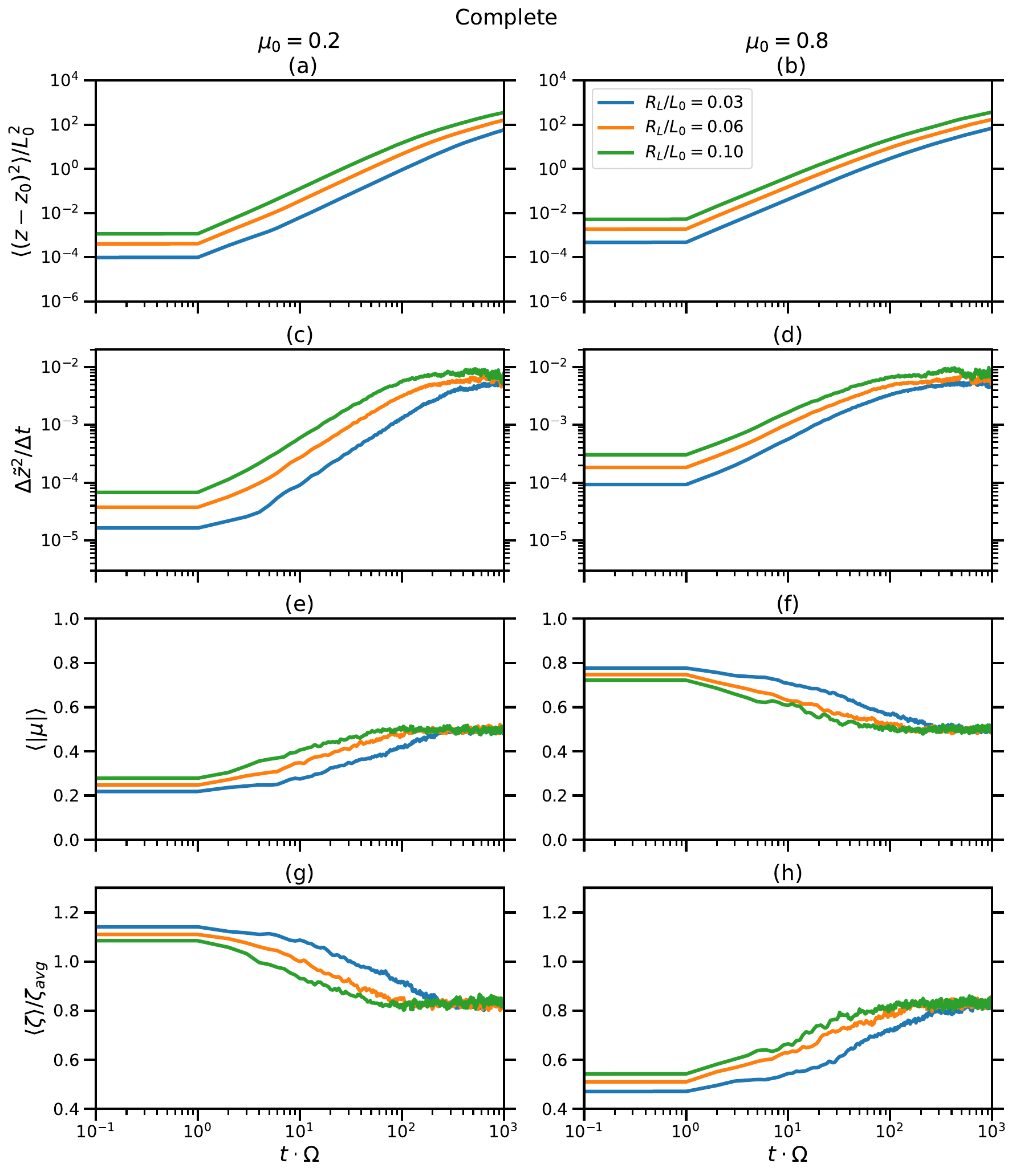}
    \caption{Top panels show the average squared displacement.
    The second row shows the  time derivative  of the top panels, $\Delta\tilde{z}^2/\Delta t=\Delta<(z-z_0)^2>/\Delta t$, giving the parallel diffusion coefficient when it becomes constant. The third row shows the average value of $\mu$ over all particles. The fourth and last row shows the average value of $\zeta/\zeta_{avg}$ (Equation \ref{eq:adiab_inv}). Different colors represent different particles Larmor radii according to the legend. Time in the x-axis is normalized by the initial value of the girofrequency $\Omega = c/R_L$ (eq. \ref{eq:larmor_radius}), with $c=1$ in code units.
   }
    \label{fig_mdzdt_complete}
\end{center}
\end{figure*}{}

\begin{figure*}[ht]
\begin{center}
 \includegraphics[width=1.9
 \columnwidth,angle=0]{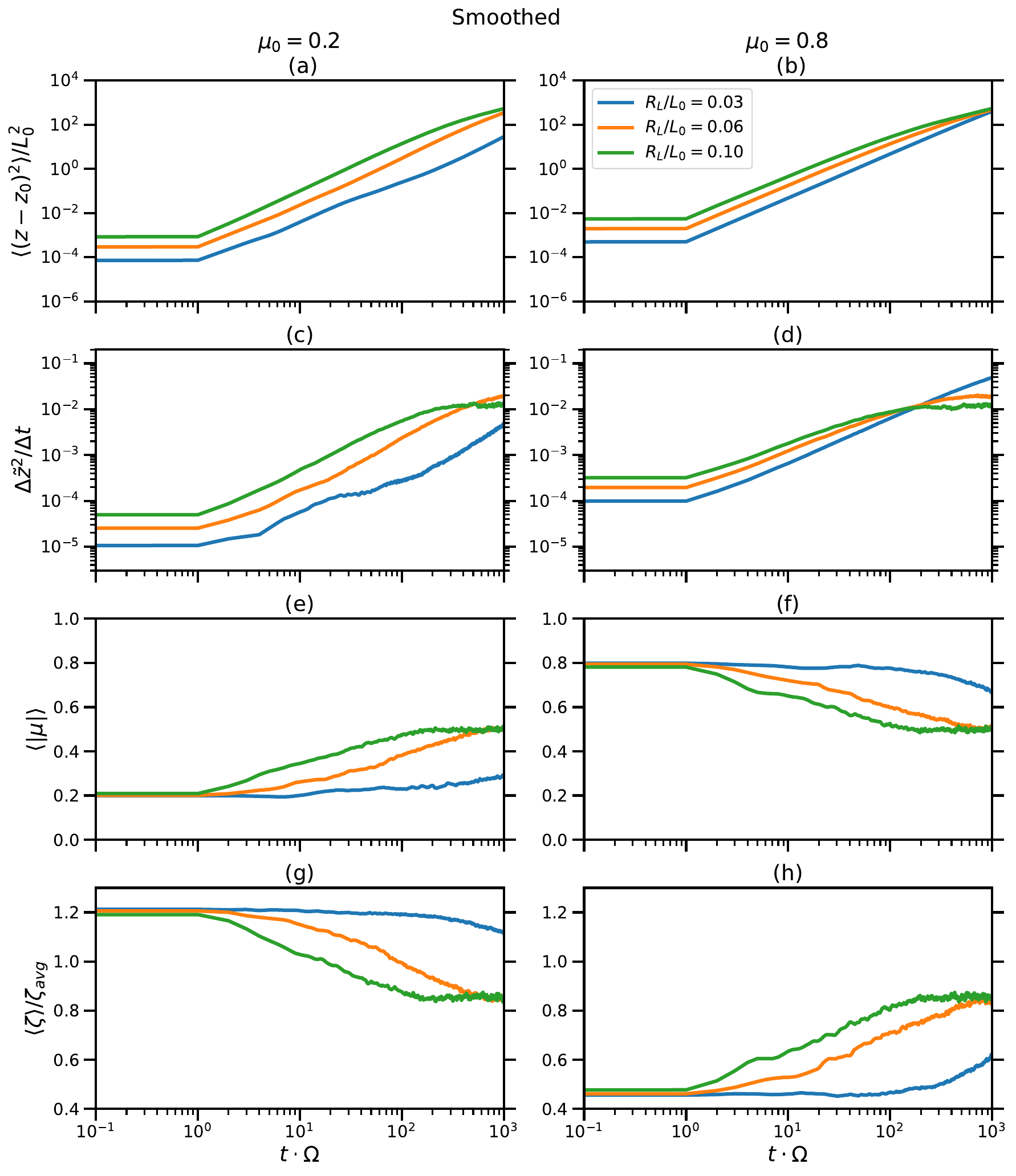}
    \caption{Same as Figure \ref{fig_mdzdt_complete}, but for models that consider the smoothed distribution of the background  magnetic field. Different colors represent different Larmor radii according to legend. 
   We note that for small Larmor radius, diffusion in $\mu$ occurs  slowly, leading to a delay in reaching the steady state within the specified time (see text for more details).}
    \label{fig_mdzdt_smoothed}
\end{center}
\end{figure*}{}

From top to bottom panels, Figure \ref{fig_mdzdt_complete} compares the squared displacement of particles, the second panel shows the evolution of first derivative of the upper panel, $\Delta\tilde{z}^2/\Delta t$ (where we write $\tilde{z}^2 = <(z-z_0)^2>$ to concise the notation), the average value of $\mu$ over all particles, and the average value of $\zeta/\zeta_{avg}$ (Equation \ref{eq:adiab_inv}) versus the normalized time $(t\times\Omega)$. The colors represent models with $R_L/L_0 = 0.03,0.06$ and $0.10$ for blue, orange and green, respectively. Note also that $\Omega$ is different for each Larmor radius. All models in this figure consider the complete magnetic field distribution of the background. 

The evolution of $\Delta\tilde{z}^2/\Delta t$ over time in Figure \ref{fig_mdzdt_complete} highlights the influence of the initial pitch angle. Initially, particles with low $\mu_0$ exhibit a smaller average squared displacement.
The case with a smaller Larmor radius requires slightly more time to reach $<\mu>\simeq0.5$. However, in all models that incorporate the full magnetic field distribution, the diffusion process stabilizes around $t\times\Omega\sim10^2$. Notably, when $\tilde{z}^2 = <(z-z_0)^2>$ increases at a slower rate and increases linearly with time, the associated diffusion coefficient, $D_\parallel= \Delta\tilde{z}^2/\Delta t$, converges to approximately $10^{-2}$ across all models.

We note that the consistent asymptotic behavior across all models, with a finite diffusion mean free path, highlights the crucial role of mirroring. The primary function of resonant scattering in this process is to introduce stochastic variations in the pitch angle, rendering diffusion eventually independent of the initial pitch angle. At the same time, mirror diffusion ensures a finite mean free path, addressing the limitation of resonant scattering, which alone fails to reverse particle moving direction—commonly known as the 90-degree problem.

Additionally, there is a dependence of the value of $D_\parallel$ on the Larmor radius, which is more apparent in Figure \ref{fig_mdzdt_complete_phys} (bottom panel, Section \ref{sec:discussion}), where a linear scale is used instead of a logarithmic one.

Figure \ref{fig_mdzdt_smoothed} depicts the same information as Figure \ref{fig_mdzdt_complete}, but for models featuring a smoothed distribution of the magnetic field as the background. The effects described earlier are more pronounced in these models. Specifically for $R_L/L_0 = 0.03$, there is a significant impact from magnetic mirrors on particles with $\mu_0 = 0.2$ as the influence of pitch-angle scattering diminishes. This is reflected in the time required for $\Delta\tilde{z}^2/\Delta t$ to stabilize. Towards the end of the observed period, $\Delta\tilde{z}^2/\Delta t$ is approximately one order of magnitude smaller for $\mu_0=0.2$ compared to $\mu_0=0.8$. 
Nevertheless, if we track the particles sufficiently long, 
this difference decreases to a factor two (see Figure \ref{fig_mdzdt_smooth_phys} where the particles have evolved for a time 2.5 times longer). Therefore, one might speculate that after a much longer evolution, $\Delta\tilde{z}^2/\Delta t$ will eventually converge to the same value for both $\mu_0$, as in the complete case.

The average behavior of the adiabatic invariant in Figure \ref{fig_mdzdt_smoothed} confirms that particles with smaller $\mu_0$ suffer mirroring for a longer duration initially. The average value of $<\zeta/\zeta_{avg}>$ for models with $\mu_0 = 0.2$ begins to exhibit more significant variations only after $t\times\Omega=10^2$, whereas for models with $\mu_0 = 0.8$, variations commence after $t\times\Omega=10^1$.

This behavior can be elucidated by considering the following. Particles marginally interact with fluctuations smaller than their Larmor radius, and pitch angle scattering is associated with interactions with fluctuations having a wavelength $\lambda\sim R_L$. At the same time, large-scale perturbations induce the formation of magnetic mirrors.
In essence, for $\mu_0=0.8$, particles can travel along magnetic field lines at much higher velocities, barely encountering mirrors initially due to their small pitch angle. 
Essentially, the initial pitch angle is sufficiently small for the particles to undergo resonant scattering with the perturbations present in the system. The condition for this interaction will be discussed in more details in the next section. This process initiates variations in $<\zeta/\zeta_{avg}>$ earlier than in  the scenario when $\mu_0=0.2$. In this latter case, particles remain confined by a series of magnetic mirrors, predominantly diffusing perpendicular to the mean magnetic field.

The effect of mirrors and scattering over the particles with smaller $R_L$ in Figures \ref{fig_mdzdt_complete} and \ref{fig_mdzdt_smoothed} is two-fold, as evidenced by the different magnetic field backgrounds. Mirrors can effectively confine particles with large pitch-angles in systems where scattering is suppressed. However, particles may escape quickly if their initial propagation direction is nearly parallel to the magnetic field, as resonant scattering then becomes significant. At the same time, resonant scattering also acts as a regulating mechanism, enabling mirror diffusion as it causes changes in the pitch angle, allowing some particles to transit to a mirror diffusion dominated regime.

An illustrative demonstration of this regulatory effect can be observed in the models depicted in Figure \ref{fig_mdzdt_smoothed}, with $R_L/L_0$ values of 0.06 and 0.10 (corresponding to $k\sim17$ and $k\sim10$, respectively). The temporal evolution of $\Delta\tilde{z}^2/\Delta t$ in these models closely resembles that of the scenario with a complete background magnetic field distribution (Figure \ref{fig_mdzdt_complete}), albeit with some nuanced distinctions. 
As the Larmor radius increases, the resonant condition is more readily satisfied, enabling more efficient  pitch-angle scattering and causing the particle behavior to more closely resemble that observed in  the complete background field case.
Models with $R_L/L_0 = 0.06$ experience less scattering compared to those with $R_L/L_0 = 0.10$, yet the former ends up with a slightly higher diffusion coefficient than the latter. This seems to suggest that scattering becomes more efficient at higher energies, but we need to take into account that the dissipative scale for this background is around $\sim0.2L_0$ (see Section \ref{sec:testpartsim}). As the Larmor radius of the models becomes close to the dissipative scale, they can be scattered more efficiently even at higher $\mu$. This can also explain why the green models ($R_L/L_0 = 0.10$) seem to be better confined than the orange ones ($R_L/L_0 = 0.06$). At a small $\mu_0$, the non-resonant mirroring can well confine particles especially when the Larmor radius is much smaller than the dissipative scale.
Though not very clear from Figure \ref{fig_mdzdt_smoothed}, it is important to note that the $R_L/L_0 = 0.03$ models (blue curves) also reach saturation at a later time. This is more evident in the bottom panels of Figure \ref{fig_mdzdt_smooth_phys}, where we have plotted a much larger time interval. Here, $<\Delta \tilde{z}^2>/\Delta t$ saturates  for both $\mu_0=0.2$ and $0.8$, with  the associated diffusion being at most one order of magnitude larger than the values reached in the complete models. Figure \ref{fig_mudist_last} illustrates the $\mu$ distribution near the end of the time interval presented in Figure \ref{fig_mdzdt_smooth_phys}. The results are shown for both smoothed models, with $\mu_0=0.2$ and $0.8$, and a Larmor radius of $R_L/L_0=0.03$. While the distribution is nearly constant and approaches isotropy, it does not become fully isotropic because of the inefficient scattering. The green curve (corresponding to $\mu_0=0.8$), for instance, shows a slightly excess of events close to $\mu=1.0$.

Finally, Figures \ref{fig_mdzdt_complete} and \ref{fig_mdzdt_smoothed} also indicate that, although the particles can achieve a close to constant $\Delta\tilde{z}^2/\Delta t$ very quickly, which is indicative of a stable diffusion process, the entire process actually occurs over 5 to 10 times the size of the simulated box. This means that this initial phase of the diffusion, which can be heavily affected by mirrors, can occur over a considerably sized region around the source. Note however that this depends on how efficient resonant scattering is and it will also depend on the particle's energy. Applications of these results and comparisons to observations will be discussed in the next section.

\section{Discussion}\label{sec:discussion}

\begin{table}[ht]
\centering
\begin{tabular}{cccccc}
\hline
$B(\mu G)$ & $L_0(pc)$ & $R/L_0$ & E(PeV) & $D_\parallel$ & $D_{\parallel,phys}(cm^2/s)$ \\ \hline
3.2        & 1         & 0.03    & 0.09   & 0.005          & $4.63\times10^{26}$            \\
3.2        & 1         & 0.06    & 0.18   & 0.006          & $5.55\times10^{26}$              \\
3.2        & 1         & 0.10    & 0.30   & 0.008          & $7.40\times10^{26}$              \\\hline
3.2        & 5         & 0.03    & 0.44   & 0.005          & $2.31\times10^{27}$              \\
3.2        & 5         & 0.06    & 0.89   & 0.006          & $2.78\times10^{27}$              \\
3.2        & 5         & 0.10    & 1.48   & 0.008          & $3.70\times10^{27}$              \\\hline
10         & 5         & 0.03    & 1.40   & 0.005          & $2.31\times10^{27}$              \\
10         & 5         & 0.06    & 2.80   & 0.006          & $2.78\times10^{27}$              \\
10         & 5         & 0.10    & 4.67   & 0.008          & $3.70\times10^{27}$              \\ \hline

\end{tabular}
\caption{Diffusion coefficients as obtained from our models with complete magnetic field distribution (Figures \ref{fig_mdzdt_complete} and \ref{fig_mdzdt_complete_phys}), for different system sizes and magnetic fields, and different particle's energy. 
}
\label{tab:phys_norm}
\end{table}

\begin{figure*}[ht]
\begin{center}
 \includegraphics[width=1.5
 \columnwidth,angle=0]{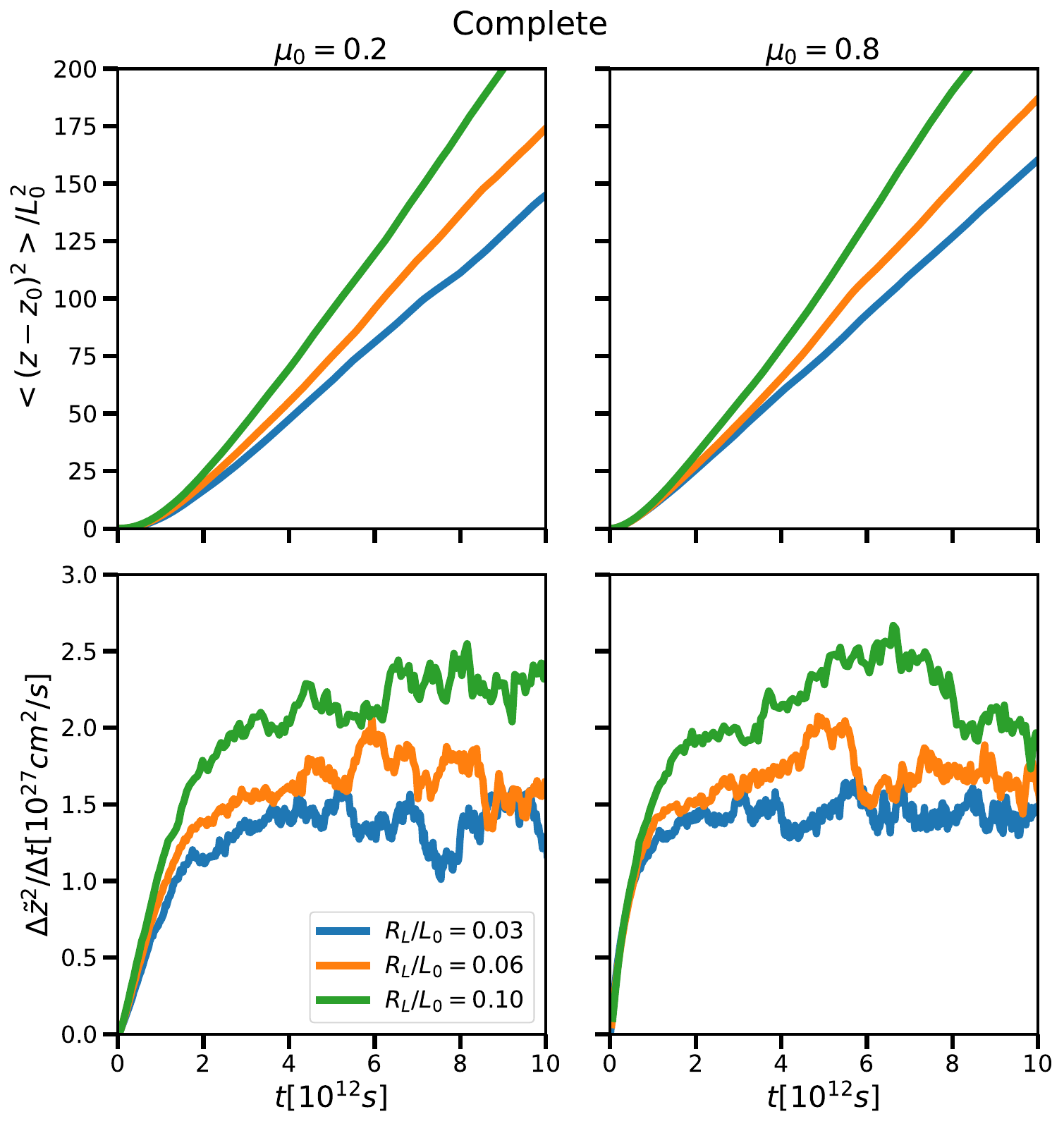}
    \caption{The same as in Figure \ref{fig_mdzdt_complete} for a complete magnetic field distribution, but considering the physical values of the particles displacement and the parallel diffusion coefficient, taking a system with size  $L_0 = 3pc$ (see Equation \ref{eq:D_phys_norm}).} 
    \label{fig_mdzdt_complete_phys}
\end{center}
\end{figure*}{}

\begin{figure*}[ht]
\begin{center}
 \includegraphics[width=1.5
 \columnwidth,angle=0]{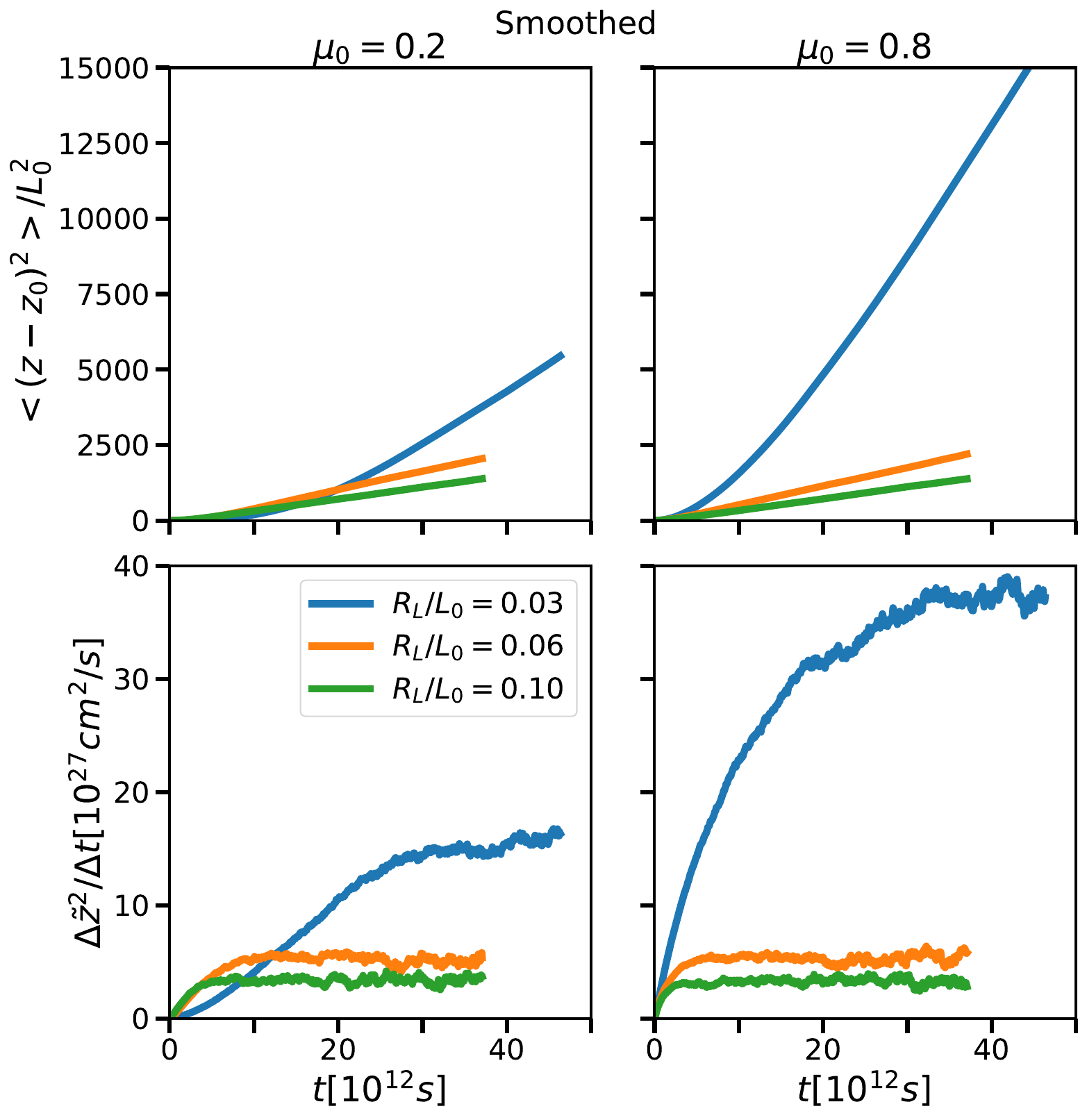}
    \caption{The same as in Figure \ref{fig_mdzdt_complete_phys}, but considering a smoothed background distribution of the magnetic field (as in Figure \ref{fig_mdzdt_smoothed}), and a much larger time interval. Models with $R_L=0.03L_0$ take a longer time to saturate due to the reduced scattering. See text for further details.}
    \label{fig_mdzdt_smooth_phys}
\end{center}
\end{figure*}{}

\begin{figure}[ht]
\begin{center}
 \includegraphics[width=0.95
 \columnwidth,angle=0]{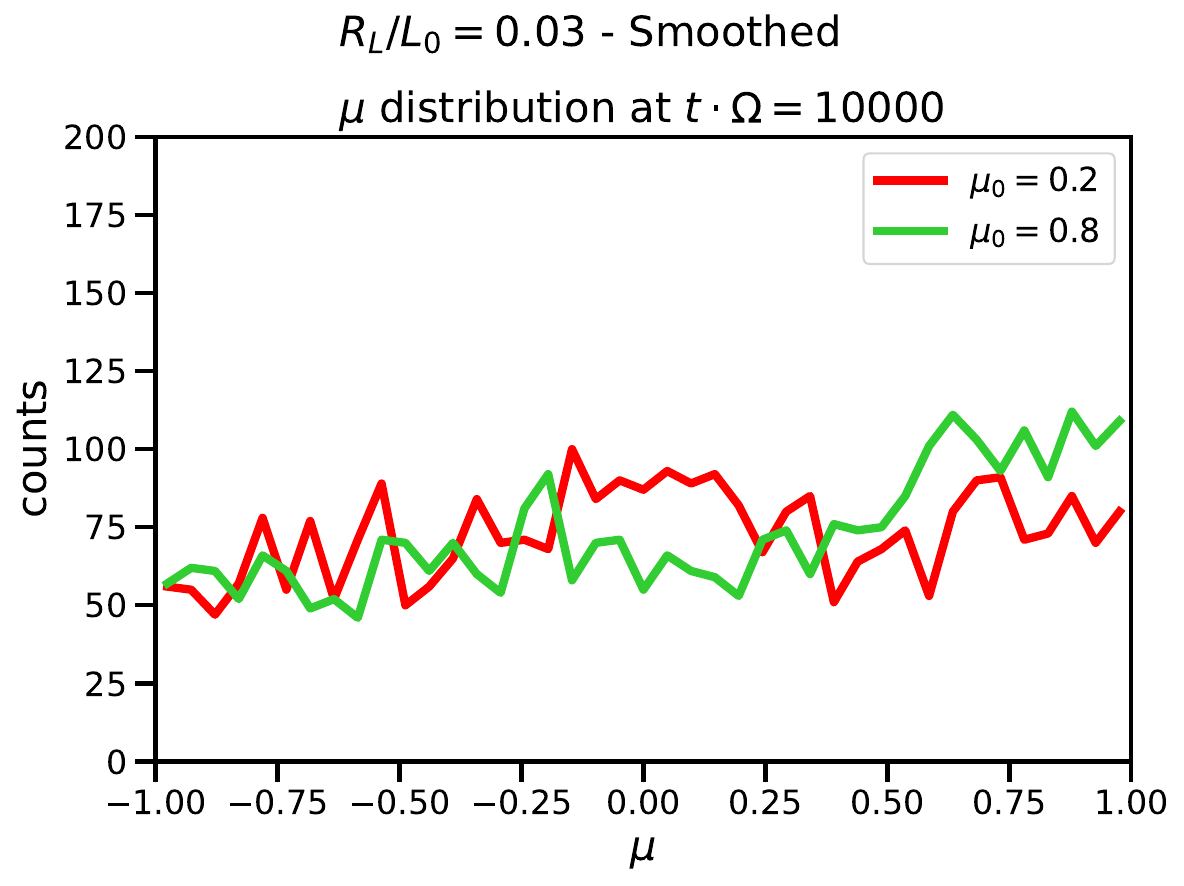}
    \caption{Distribution of $\mu=\cos{\theta}$ at $t\cdot\Omega=10^4$ (or $t=4.6\cdot10^{13}s$) for the smoothed models with $R_L/L_0=0.03$ (blue lines in Figure \ref{fig_mdzdt_smooth_phys}).
    } 
    \label{fig_mudist_last}
\end{center}
\end{figure}{}

In this work we have explored the effects of mirror diffusion of CRs by combining 3D MHD simulations of a supersonic, sub-Alfvenic turbulent  molecular cloud with test particle simulations, considering initial pitch angles  corresponding to $\mu = 0.2$ and $0.8$.
We considered two background magnetic field distributions - the smoothed and the complete models - for  CR propagation analysis. The study of the smoothed model was entailed only to highlight the effects of mirror diffusion over scattering. Nevertheless, in realistic astrophysical media, the damping scale of MHD turbulence can be larger or smaller than the gyroradius of the CRs under consideration. Therefore, both models investigated (smoothed and complete) are realistic.

Particles injected into the turbulent domain undergo both bouncing (or mirroring)  and resonant scattering processes. We have observed that mirror diffusion significantly influences their parallel diffusion component relative to the mean magnetic field, which is consistent with LX21 predictions. Turbulent magnetic mirrors can still happen.

Analyzing their $\mu_b$, defined as the last maximum absolute value of $\mu$ where their adiabatic invariant $\zeta$ exhibits minimal variation over time, reveals that the majority of bounces occur for $\mu_b<0.6$, however, this should depend on the background magnetic field.

The initial diffusion of CRs is influenced by their initial pitch angle, with those starting closer to $90^\circ$ (or $\mu_0$ closer to 0) being more prone to mirror diffusion and requiring longer diffusion times, while those with smaller pitch angles (larger $\mu_0$) can escape the system more rapidly. This disparity also affects the time required for particles to attain normal diffusion. Particles beginning with pitch angles near $90^\circ$ achieve a nearly constant $\Delta\tilde{z}^2/\Delta t$ at a slower rate (see Figures \ref{fig_mdzdt_complete} and \ref{fig_mdzdt_smoothed}). Suppression of pitch angle scattering can further accentuate the role of the initial pitch angle in defining the particle diffusion regime, depending on its Larmor radius.

We can convert the parallel diffusion coefficient $D_\parallel$ (defined as the value of $\Delta\tilde{z}^2/\Delta t$ after it stabilizes) obtained from the simulations in code units into physical units, by considering an arbitrary physical size for the system, as follows:

\begin{equation}
\begin{split}
    D_{\parallel,phys}[\mathrm{cm^2/s}] &= D_\parallel \cdot L_0\,c\\
    &=D_\parallel\cdot2.776\times10^{29}\,\Big(\frac{L_0}{3\,\mathrm{pc}}\Big),\label{eq:D_phys_norm}
\end{split}
\end{equation}

\noindent where $L_0$ denotes the physical dimension of the system under examination, and $c$ is the light speed.

We can also convert the magnetic field from our simulations in code units into physical units, as follows:
\begin{equation}
\begin{split}
    B[\mathrm{\mu G}]=3.215\,\Big(\frac{n}{10\,\mathrm{cm^{-3}}}\Big)^{\frac{1}{2}}\Big(\frac{c_s}{1.9\cdot10^4\,\mathrm{cm/s}}\Big)\\
    \Big(\frac{M_s}{7.0}\Big)\Big(\frac{M_A}{0.6}\Big)^{-1},\label{eq:B_phys_norm}
\end{split}
\end{equation}

\noindent where $n$ is the number density, and $c_s$ is the sound speed of the system. The values adopted in this equation  are representative of  a molecular cloud region, but in Table \ref{tab:phys_norm} we also consider  parameters from other systems.

Considering that particles travel close to speed of light, we can write:

\begin{eqnarray}
    R_L[\mathrm{pc}] &=& 1.084 \frac{(E/1\mathrm{PeV})}{(B/1\mathrm{\mu G})}\Big(\frac{v_\perp}{c}\Big),\label{eq:RL_phys}\\
    \Omega[\mathrm{s^{-1}}] &\approx& 9.743\times10^{-9}\Big(\frac{1\mathrm{pc}}{R_L}\Big),\label{eq:Omega_phys}
\end{eqnarray}

\noindent where $E$ is the energy of the particle, $\Omega$ its gyrofrequency, and $R_L$ its Larmor radius in second and pc units, respectively. As stressed before, we have assumed $v_\perp \approx c$.

As emphasized in Section \ref{sec:intro}, recent findings from VHE gamma-ray observatories like H.E.S.S., HAWC and LHAASO have revealed extensive diffuse emissions reaching a few to hundreds of TeV and even PeV within our galaxy \citep{2017Sci...358..911A,2018ApJ...866..143H,2021NatAs...5..465A,2021Natur.594...33C}, without definitive identification of their sources. Consequently, comprehending CR acceleration and diffusion in the surroundings of embedded  sources  has become critically important for interpreting the observations. Traditional models and assumptions to characterize CR diffusion are insufficient to elucidate these emissions.

To compare our results with  observations, we present in Figure \ref{fig_mdzdt_complete_phys} the same diagrams as in Figure \ref{fig_mdzdt_complete} for the complete magnetic field distribution, but depicting the physical parallel diffusion coefficient, $(\Delta\tilde{z}^2/\Delta t)|_{phys}$ 
(see Equation \ref{eq:D_phys_norm}), as a function of the time ($L_0/c$) in seconds. For the physical size of the  system, we have adopted $L_0= 3$pc in this figure.

Combining the results of Figures \ref{fig_mdzdt_complete} and \ref{fig_mdzdt_complete_phys}, with Equation \ref{eq:RL_phys} of the Larmor radius to evaluate the particle's energy for a given background magnetic field, we can obtain the corresponding parallel diffusion coefficient for the particle, as predicted from our models. Table \ref{tab:phys_norm} shows $D_{\parallel,phys}$ for particles with different energies and magnetic fields in systems with different sizes. We also observe that particles travel approximately 10 times the scale $L_0$ before attaining the saturation point (i.e., a  nearly constant value) of the parallel diffusion coefficient. Throughout the initial stage, the diffusion of the particles varies based on their initial pitch angle, with greater susceptibility to mirror effects when the pitch angle approaches $90^\circ$.

We note that $D_{\parallel,phys}$ is consistent with the implied values to explain VHE observations of extended sources like Geminga and PSR B0656+14 \citep{2017Sci...358..911A}. Their combined halo has an inferred diffusion coefficient of $4.5\times10^{27}\mathrm{cm^2/s}$ for 100 TeV CRs within an extended zone with tens of parsec \citep{2017Sci...358..911A}. Similarly, \cite{2018ApJ...866..143H} propose a diffusion coefficient $\lesssim10^{28}\,\mathrm{cm^2/s}$ for electrons of $\sim10$ TeV in a region of a few tens of parsecs around Vela X. \cite{2022FrASS...922100F} provides a comprehensive review of the characteristics of recent PWNe VHE observations and possible scenarios to explain the small diffusion coefficients which are also compatible with our findings. We refer to his work for more detailed information.

Interestingly, our results from the models featuring smoothed background magnetic field distribution (Figures \ref{fig_mdzdt_smoothed} and \ref{fig_mdzdt_smooth_phys}) could also offer insights into systems like the nuclear regions of starburst galaxies. For example, \cite{2020MNRAS.493.2817K} argues that, as in molecular clouds within our Galaxy, these systems are likely to possess sufficiently low ionization fractions. Consequently, ion-neutral damping would halt turbulent cascades at scales much larger than the gyroradius of CRs with energies in the range of a few hundred TeV. Still, CRs could scatter off turbulence self-generated by streaming instability. The competition between the streaming instability and ion-neutral damping leads to a CR streaming speed equal to the ion Alfvén speed, which is independent of CR energy up to $\sim 1 \mathrm{TeV}$. Diffusion should occur only due to field line random walk along the magnetic field, yielding a diffusion coefficient of approximately $10^{27}-10^{28} \,\mathrm{cm^2/s}$, in this energy range.
While the behavior of these CRs would align with the results shown in Figures \ref{fig_mdzdt_smoothed} and \ref{fig_mdzdt_smooth_phys} in the absence of streaming instability feedback, proper modeling of diffusion in such environments necessitates the incorporation of CR feedback effects too.

Resonant scattering occurs only if the resonance condition is satisfied. For the resonant wavenumber $k_{res}$, we can write:

\begin{align}
    k_{res} 2 \pi R_L \simeq \frac{(1-\cos^2{\theta_{res}})^{0.5}}{\cos{\theta_{res}}}, 
\end{align}

\noindent where $\theta_{res}$ is the angle between the particle’s velocity and the magnetic field that satisfies the resonance condition. In our smoothed models, smoothing limits $k_{res}$ to values below 5. For $R_L = 0.03L_0$, we find $\tan{\theta_{res}} \sim 0.9425$ or $\mu_{res} \sim 0.73$ ($\theta_{res} \sim 43^\circ$). This implies that resonant scattering 
will be dominant only for approximately $\mu \gtrsim 0.73$. Referring to Figure \ref{fig_mucXmub}, all bounces occur with $\mu_{c,\text{max}} < 0.7$ (panels (c) and (d)). However, some particles eventually cease bouncing and align more closely with the magnetic field, reaching $\mu \gtrsim 0.73$. This transition triggers resonant scattering, enabling particles to diffuse more quickly across the system (see Figure \ref{fig_testPart_muXt_r15_k5}, see also Appendix \ref{appendix:scattering_displacement}). This mechanism explains the higher diffusion coefficient observed in the smoothed case compared to the complete magnetic field distribution for $R_L = 0.03L_0$ (Figures \ref{fig_mdzdt_complete} and \ref{fig_mdzdt_complete_phys}).

Indeed, in the complete case, mirroring remains dominant for $\mu \lesssim 0.7$, but scattering occurs across all $\mu$ values. Consequently, in a fully developed magnetic field, the interplay between these mechanisms at large pitch angles becomes more frequent starting already at small spatial scales, leading to more transitions. These transitions slow diffusion significantly compared to the smoothed scenario.

This behavior is evident in all models with $R_L/L_0 = 0.03$ (indicated by blue curves). Comparing diffusion coefficients in Figures \ref{fig_mdzdt_complete_phys} and \ref{fig_mdzdt_smooth_phys}, we observe that in the complete magnetic field distribution, $D_\parallel$ stabilizes around $\sim 10^{27} \text{cm}^2/\text{s}$, whereas in the smoothed case, $D_\parallel$ continues to increase by nearly an order of magnitude more,  with a distinct trend of delayed saturation.

For $R_L = 0.06L_0$, the resonant condition gives $\theta_{res} \lesssim 63^\circ$ or $\mu_{res} \gtrsim 0.46$. For $R_L = 0.10L_0$, $\theta_{res} \lesssim 73^\circ$ or $\mu_{res} \gtrsim 0.30$. This means that for $R_L = 0.03L_0$, the range of $\mu$ values subject to scattering is narrower. Particles with larger $R_L$ transition more readily between the two regimes, explaining their behavior, which resembles that of the complete magnetic field distribution models. While both scattering and mirroring are present in all cases, $R_L = 0.03L_0$ provides the clearest distinction between these processes.


The discussion above can be further stressed.
It is important to note that suppressing small-scale structures in the smoothed models (Figures \ref{fig_mdzdt_smoothed} and \ref{fig_mdzdt_smooth_phys}) not only reduces the effectiveness of pitch-angle scattering but also eliminates small-scale magnetic mirrors, as discussed in Section \ref{sec:ang_distribution}. These small-scale mirrors are more effective at reflecting particles due to their stronger longitudinal magnetic field gradients compared to larger-scale mirrors. In Figures  \ref{fig_mdzdt_smoothed} and \ref{fig_mdzdt_smooth_phys}, the scattering is reduced  for the blue curve. At early times, diffusion with $\mu_0 = 0.2$ is slower than that with $\mu_0 = 0.8$ due to the stronger mirroring at small $\mu$. Later in Figure \ref{fig_mdzdt_smooth_phys}, as some particles experience an increase in $\mu$—despite inefficient scattering—the reduced scattering at large $\mu$ accelerates overall diffusion, resulting in faster transport. Thus, in the smoothed magnetic field, diffusion becomes more efficient compared to the complete field case, reinforcing the conclusions presented earlier.


The mean free path in scattering regime alone at small pitch angles would be infinity. However, in reality, it is not possible to study scattering regime in isolation, as scattering inevitably causes particles to transit between scattering and mirroring regimes - i.e., small to large pitch angles (and vice versa). Our study with a smoothed magnetic field (which still includes  weak scattering) clearly shows this trend (see the blue line in Figure \ref{fig_mdzdt_smooth_phys}). In the case of mirroring, it is possible to separately study its effect on diffusion near the source of CRs, as scattering (small pitch-angle) CRs escape rapidly, leaving  behind only mirroring (large pitch-angle) CRs near the source. This entails an enhancement of CRs, due to their slow mirror diffusion, and gamma rays specially near the source, an effect that is observable \citep[e.g.][]{2017Sci...358..911A}.


\cite{2023ApJ...959L...8Z} have recently focused on the mirror diffusion alone, testing on the pitch-angle-dependent mean free path of mirror diffusion. Here we have studied the overall propagation behavior with both mirror
and scattering diffusion taken into account. We find that the propagation behavior in the early time, e.g., in the vicinity of CR sources, depends on the initial pitch angle distribution. Away from CR sources, it transits from the initial parallel superdiffusion to the parallel normal diffusion. 

It is worth noting that recent studies by \cite{2023JPlPh..89e1701L} and \cite{2023MNRAS.525.4985K} have examined the implications of intermittent bends and reversals in magnetic fields on CR diffusion. They conclude that low-energy particles experience more efficient confinement than high-energy particles, attributed to the absence of pitch angle scattering and isotropization. All their models consider very small Larmor radius, and  the only model among ours that has a similar initial configuration for the particles is the one with a smoothed background magnetic field with $R_L/L_0=0.03$. However, while we consider supersonic and sub-Alfvénic turbulence, those works employ different  background magnetic field configurations. 
The reduced influence of pitch angle scattering is a significant factor influencing fast particle escape rates ($\mu_0=0.8$) or prolonged mirroring ($\mu_0=0.2$). However, further investigation is warranted to explore the intersections between their studies and ours. More recently, \cite{2024arXiv240603542Z} have investigated the influence of mirror diffusion in a similar scenario to that employed by \cite{2023JPlPh..89e1701L} and \cite{2023MNRAS.525.4985K}. We refer to their work for a comparison between these different processes.

Lastly, a recent study by \cite{2025MNRAS.539.1236B} demonstrated that magnetic mirroring, combined with modest small-angle scattering, can reduce cosmic ray transport to levels approaching Bohm diffusion. Their Vlasov-Fokker-Planck analysis shows how large-scale magnetic mirrors effectively confine CRs near shocks, potentially enabling acceleration to PeV energies without the need for magnetic field amplification on the Larmor scale. By bridging the gap between idealized diffusion models and realistic magnetic field geometries, their work—complementary to the present study—reinforces the need to incorporate mirroring effects in theories of CR acceleration and propagation.

\section{Conclusions}\label{sec:conclusion}

In this work, we have numerically demonstrated the process of mirror diffusion theoretically predicted in LX21. This effect is particularly significant for particles with large pitch angles. We found that in combination with resonance scattering, the particle transport is akin to a Levy-flight-like propagation. Our numerical results agree with those in \cite{2023ApJ...959L...8Z}, but extended them accounting for both mirror and resonant scattering diffusion to assess their combined effects aiming at applications to realistic astrophysical conditions.

Mirror effects are essential for small $\mu$ propagation (large pitch angle), and so they are relevant while addressing the $\sim 90-$degree scattering problem and its impact on the effective diffusion.

We have found that when pitch angle resonant scattering is significantly suppressed, the impact of mirror diffusion depends on the particle's initial pitch angle, for a significant amount of time. This effect is especially evident for lower energy particles.

Furthermore, our findings have revealed that the range of pitch angles subjective to mirroring aligns with the predictions derived from the models of \cite{2020ApJ...894...63X} and LX21, and the estimated physical values of the diffusion coefficient, $\lesssim 10^{27} \mathrm{cm^2/s}$, fall within recent constraints imposed by VHE energy observations. 

Resonant scattering at small pitch angles alone would lead to an infinite particle mean free path, as it  cannot reverse the direction of particle motion.
The combination of resonance scattering with mirror diffusion leads to a slower diffusion compared with predictions that consider
scattering alone of particles \citep[see for a discussion e.g.][and references therein]{2001ApJ...553..198F,2009A&A...507..589S,2010A&A...524A..51D,2016A&A...588A..73C,2017Sci...358..911A,2020MNRAS.493.2817K},
and we contend that mirror diffusion should be considered when studying CR propagation within turbulent magnetized environments. Moving forward, we intend to integrate these effects into comprehensive simulations of CR propagation in massive star-forming regions in order to quantify their VHE emission.


\begin{acknowledgments}
L.B-M. acknowledges support from the Brazilian funding agency CAPES, CNPq and also the kind hospitality of Professor Alex Lazarian and the University of Wisconsin - Madison; E.M.G.D.P. acknowledges support from the Brazilian funding agencies FAPESP (grants 2013/10559-5 and 2021/02120-0) and CNPq (grant 308643/2017-8). SX acknowledges the support from NASA ATP award 80NSSC24K0896. A.L. acknowledges the support of 823 wNSF grants AST 2307840.
This research was also supported in part by grant no. NSF PHY-2309135 to the Kavli Institute for Theoretical Physics (KITP), and by 
 the Aspen Center for Physics through the NSF grant PHY-2210452 and Durand Fund.
The simulations presented in this work were performed in the cluster of the Group of Plasmas and High-Energy Astrophysics of IAG-USP (GAPAE) acquired with support from FAPESP (grant 2013/10559-5). We also acknowledge  inspiring and very fruitful discussions with Anthony Bell, Philipp Kempski, Kun Fang and the participants of the program on “Turbulence in Astrophysical Environments” at KITP. 
\end{acknowledgments}

\section*{Data Availability}
The data of this article will be available upon request to the corresponding author.

\bibliography{sample631}{}
\bibliographystyle{aasjournal}

\appendix
\section{Pitch angle scattering and large particle displacement}\label{appendix:scattering_displacement}

\begin{figure}[ht]
\begin{center}
 \includegraphics[width=0.43
 \columnwidth,angle=0]{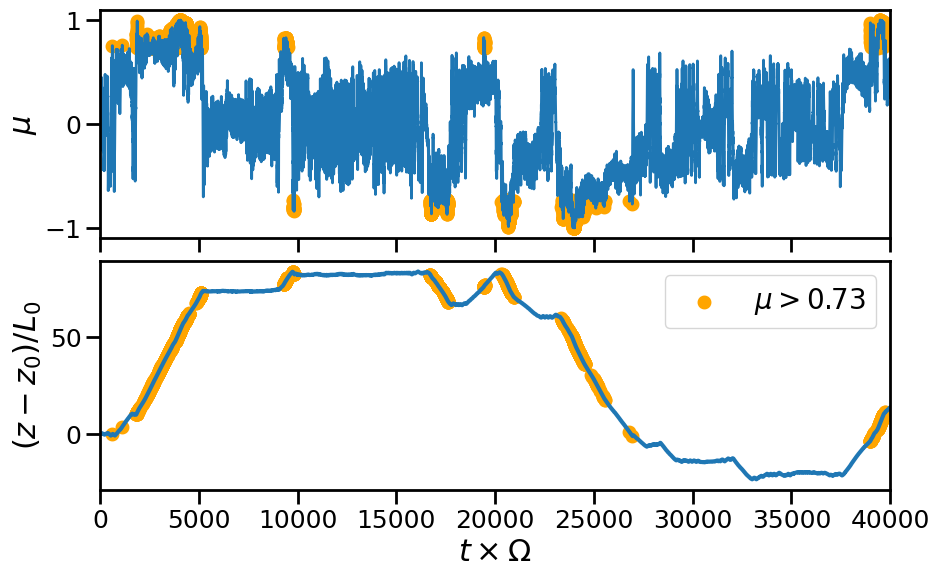}
    \caption{Particle's pitch-angle evolution and displacment parallel to the magnetic field along time. Initially the particle has $\mu_0=0.2$, $R_L/L_0=0.03$, and a smoothed background magnetic field distribution. Orange regions around the blue line show parts of the trajectory where the estimated condition for scattering is satisfied.
    } 
    \label{fig_scattering-displacement}
\end{center}
\end{figure}{}

Figure \ref{fig_scattering-displacement} illustrates the trajectory of a particle characterized by $\mu_0 = 0.2$, $R_L = 0.03L_0$, within a smoothed background magnetic field. The orange circles highlight segments of the trajectory where $\mu > 0.73$, which, as discussed in Section \ref{sec:discussion}, corresponds to the scattering regime for particles in this type of magnetic field background. This example demonstrates that the particle exhibits the most significant diffusion across the system when it is in the scattering regime.

\section{Alternative method to identify bounces}\label{appendix:alt_mirror_criteria}

\begin{figure}[ht]
\begin{center}
 \includegraphics[width=0.45
 \columnwidth,angle=0]{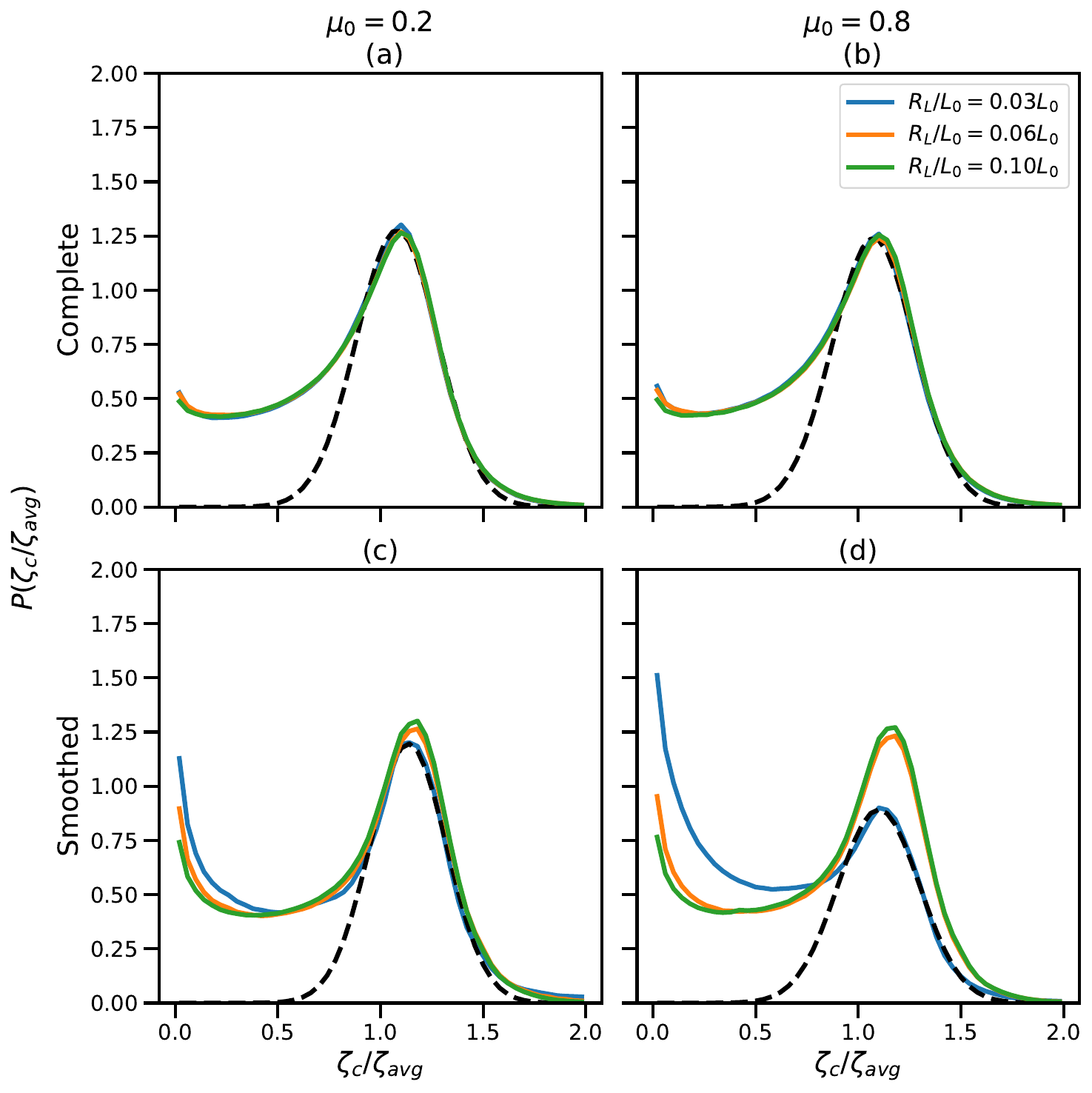}
    \caption{Distribution of $\zeta_c/\zeta_{avg}$ for all bounce candidates in each simulation. The black dashed line shows 
    a Gaussian distribution  fit to the characteristic peak around $\zeta_c/\zeta_{avg}=1$ for models with $R_L = 0.03L_0$.
    } 
    \label{fig_zetab_dist}
\end{center}
\end{figure}{}

\begin{figure}[ht]
\begin{center}
 \includegraphics[width=0.50
 \columnwidth,angle=0]{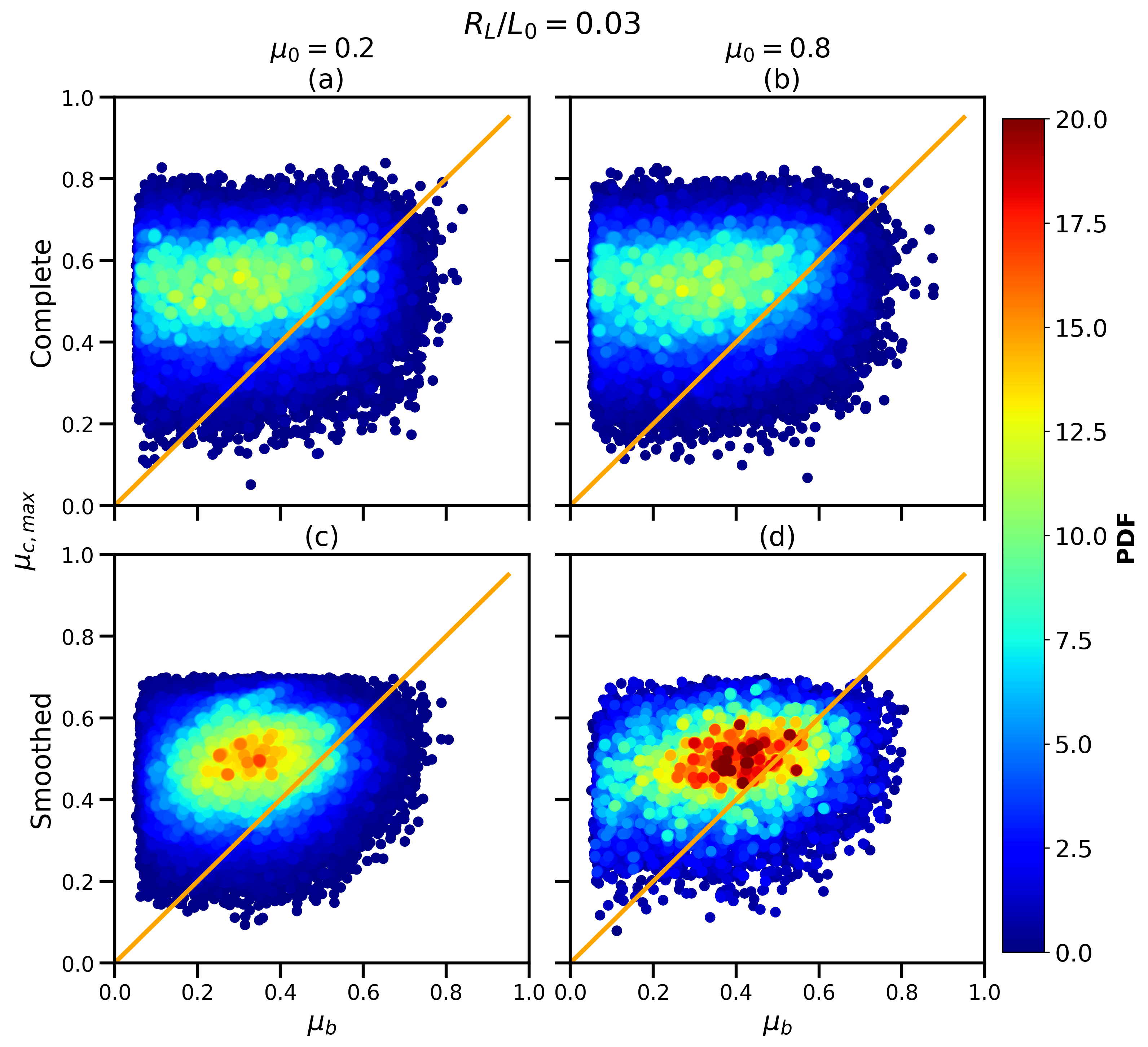}
 \includegraphics[width=0.445
 \columnwidth,angle=0]{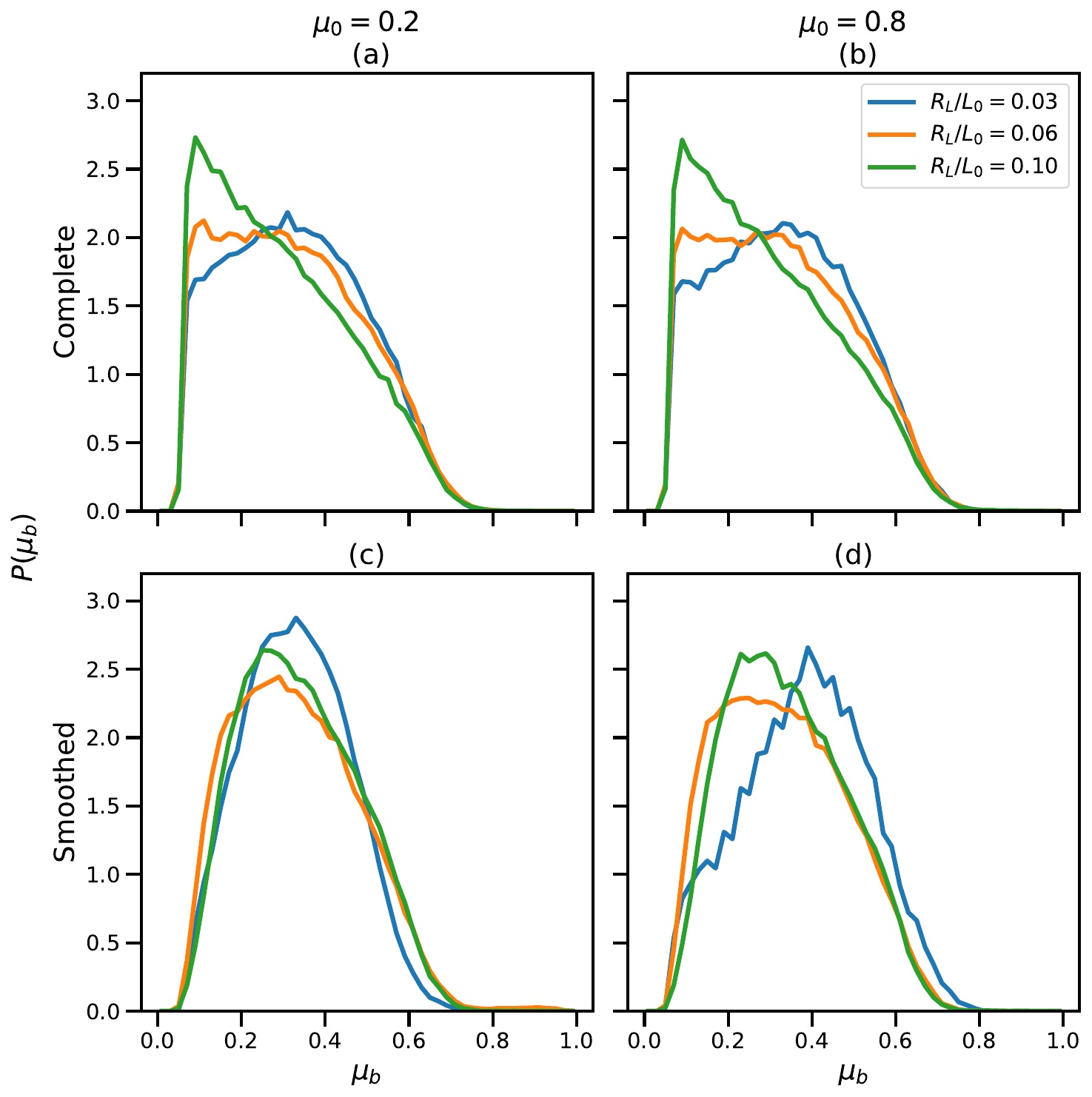}
    \caption{Same as Figures \ref{fig_mucXmub} and \ref{fig_PDF_bounceMu} but using Gaussian method for the identification of $\mu_b$.
    } 
    \label{fig_appendix_distributions}
\end{center}
\end{figure}{}

\begin{table}[]
\centering
\caption{Same as Table \ref{tab:mirror-events} but using the Gaussian method to identify mirror bounces.}
\label{tab:mirror-events-gauss}
\begin{tabular}{cccccccc}
\hline
Complete & $\mu_0=0.2$   &        &       &  & $\mu_0=0.8$   &        &       \\ \hline
$R_L$    & Mirror Bounce & Total  & \%    &  & Mirror Bounce & Total  & \%    \\ \hline
0.03     & 91361         & 123859 & 73.76 &  & 89931         & 120331 & 74.74 \\ 
0.06     & 165459        & 223060 & 74.18 &  & 161250        & 218128 & 73.92 \\ 
0.10     & 246912        & 322056 & 76.67 &  & 234385        & 307357 & 76.26 \\ \hline\hline
Smooth   & $\mu_0=0.2$   &        &       &  & $\mu_0=0.8$   &        &       \\ \hline
$R_L$    & Mirror Bounce & Total  & \%    &  & Mirror Bounce & Total  & \%    \\ \hline
0.03     & 247899        & 354073 & 70.01 &  & 110200        & 152117 & 72.44 \\ 
0.06     & 126401        & 154984 & 81.56 &  & 114512        & 136214 & 84.07 \\ 
0.10     & 108227        & 140372 & 77.10 &  & 105874        & 137993 & 76.72 \\ \hline
\end{tabular}
\end{table}

We propose here an alternative method to identify  bounces
to the one introduced in Section \ref{sec:trajectory_bounceid}.
We first analyze the distribution of $\zeta_c/\zeta_{avg}$ for all bounce candidates. As illustrated in Figure \ref{fig_zetab_dist}, a 
peak emerges near $\zeta_c/\zeta_{avg}=1$, coinciding with the $\zeta/\zeta_{avg}$ values where the majority of bounces occur. We then fit a Gaussian distribution to the peak and classify all values within $1.8\sigma_{fit}$ of the peak center as mirror bounces.

The results obtained using this method are largely consistent with those presented in Section \ref{sec:ang_distribution}. However, some differences also arise with regard to  Figures \ref{fig_mucXmub} and \ref{fig_PDF_bounceMu}. With this new approach, most of the data points below the orange line in Figure \ref{fig_mucXmub} are eliminated, while in Figure \ref{fig_PDF_bounceMu}, the tail of each distribution decays to zero for $\mu_b>0.8$. See Figure \ref{fig_appendix_distributions} for comparison.

Although this method  seems to  improve the results, there is still  a   selection effect. We cannot guarantee that changes in direction which are not associated to the peak around $\zeta_c/\zeta_{avg}=1$, are not due to mirrors. In other words, we could have mirror events with smaller values of $\zeta_c/\zeta_{avg}$.


Panel (d) of Figure \ref{fig_zetab_dist} highlights another important feature: a distinct secondary peak near $\zeta_c/\zeta_{avg} = 0$ for $R_L = 0.03L_0$. While this might initially suggest an increased role of scattering—since $\zeta$ approaches zero as $\mu \rightarrow 1$—it is important to consider the nature of particle motion. In the absence of significant scattering, particles move ballistically along magnetic field lines with conserved magnetic moment, a process known as wandering diffusion \citep[e.g.,][]{2024arXiv240603542Z}.
To differentiate between wandering diffusion and mirroring diffusion: the former occurs at large, nearly constant $\mu$ values (indicating no change in direction relative to the local magnetic field), whereas the latter involves small, oscillating $\mu$ with frequent crossings through $\mu = 0$.
A closer analysis shows that the Gaussian distribution accounts for over 72\% of all candidate bounces, pointing to mirroring as the dominant mechanism. Even though the selected conditions in this case favor wandering diffusion, mirroring still emerges as the primary cause of particle deceleration. In all other cases examined, mirror-induced bounces are even more clearly dominant.


Table \ref{tab:mirror-events-gauss} compares the total number of identified events. While the percentage of mirror events improves relative to Table \ref{tab:mirror-events} in most cases, the true value likely lies between the two estimates, considering the discussion above. Also, the slightly smaller percentages for smaller Larmor radius rather than larger ones, are directly related to the secondary peaks in Figure \ref{fig_zetab_dist}. Nevertheless, the analysis consistently supports the conclusion that mirroring is the dominant mechanism in CR diffusion.



\end{document}